\documentclass{aa} 
\pdfoutput=1

\usepackage{enumitem}
\setlist{parsep=0pt,listparindent=\parindent}
\usepackage{amssymb}
\usepackage{graphicx}
\usepackage{txfonts}
\usepackage{color}
\usepackage{hyperref}

\newcommand{\sIII}[1]{{\leavevmode\color{black}#1}}
\newcommand{\arif}[1]{{\leavevmode\color{black}#1}}

\newcommand{\sI}[1]{{\leavevmode\color{black}#1}}
\newcommand{\arifI}[1]{{\leavevmode\color{black}#1}}
\newcommand{\arifII}[1]{{\leavevmode\color{black}#1}}
\newcommand{\arifIII}[1]{{\leavevmode\color{black}{#1}}}
\newcommand{\LEt}[1]{{\leavevmode\color{black}{}}}

\newcommand{\be}{\begin{equation}}
\newcommand{\ee}{\end{equation}}

\newcommand{\msun}{${\rm M}_{\odot}$\,}

\begin{document} 

  \title{Feedback from reorienting AGN jets}
  \subtitle{I. Jet-ICM coupling, cavity properties and global energetics}
  \author{
    S.~Cielo\inst{1},
    A.~Babul\inst{2,3,4},
    V.~Antonuccio-Delogu\inst{5},
    J.~Silk\inst{1,4,6,7,8} \and
    M.~Volonteri\inst{1,4}
  }
  \institute{
  Sorbonne Universités, UPMC Univ Paris 6 et CNRS, UMR 7095, IAP, Paris, 98 bis bd Arago, 75014 Paris, France \\ \email{[cielo; silk; volonteri]@iap.fr} 
\and
  University of Victoria, 3800 Finnerty Road, Victoria BC V8P 5C2, Canada \\ \email{babul@uvic.ca} 
\and
  Institute of Computational Science, Centre for Theoretical Astrophysics and Cosmology, University of Zurich, Winterthurerstrasse 190, 8057, Zurich, Switzerland
\and
 Institut d'Astrophysique de Paris, 98 bis bd Arago, F-75014 Paris, France
\and
  INAF/Istituto Nazionale di Astrofisica-Catania Astrophysical Observatory, Via S. Sofia 78, I-95126 Catania, Italy \\ \email{Vincenzo.Antonuccio@oact.inaf.it}
 \and
 AIM-Paris-Saclay, CEA/DSM/IRFU, CNRS, Univ Paris 7, F-91191 Gif-sur-Yvette, France
\and
 Department of Physics and Astronomy, The Johns Hopkins University, Baltimore, MD 21218, USA
 \and
 BIPAC, University of Oxford, 1 Keble Road, Oxford OX1 3RH, UK 
  }
   \date{Received September 15, 20XX; accepted September 16, 20XX}
   \abstract
 {}
   {We test the effects of re-orienting jets from an active galactic nucleus (AGN) on the intracluster medium in a \arifII{galaxy cluster environment with short central cooling time.} We investigate both the appearance \arifII{and the properties of the resulting cavities}, and the \arifII{efficiency of the jets in providing near-isotropic heating to the cooling cluster core.} 
   }
   {\arifII{We use numerical simulations to explore \sI{four} models of AGN jets over several active/inactive cycles.   We keep the jet power and duration fixed across the models, varying only the jet re-orientation angle prescription.  We track the total energy of the intracluster medium (ICM) in the cluster core over time, and the fraction of the jet energy transferred to the ICM.   We pay particular attention to where the energy is deposited.  We also generate synthetic X-ray images of the simulated cluster and compare them qualitatively to actual observations.}
}   
   {Jets whose re-orientation is \arifII{minimal ($\lesssim 20^{\circ}$) typically produce conical structures of interconnected cavities, with the opening angle of the cones being $\sim 15-20^{\circ}$, extending to $\sim 300$ kpc from the cluster centre.  \sI{Such} jets transfer \sI{about} $60\%$ of their energy to the ICM, \sI{yet} they are not very efficient at heating the cluster core, \textcolor[rgb]{0.984314,0.00784314,0.027451}{\textcolor[rgb]{0,0,0}{and even less efficient at heating it isotropically}}, because the jet energy is deposited further out.  Jets that re-orientate by $\gtrsim 20^{\circ}$ generally produce multiple pairs of detached cavities.  \sI{A}lthough smaller, the\sI{se} cavities are inflated within the central $50$~kpc and are more isotropically distributed, resulting in more effective heating of the core.  Such jets, over hundreds of millions of years, can deposit up to $80\%$ of their energy precisely where it is required.  Consequently, these models come the closest in terms of approaching a heating/cooling balance and mitigating runaway cooling of the cluster core even though all models have identical jet power/duration profiles.  Additionally, the corresponding synthetic X-ray images exhibit structures and features closely resembling those seen in real cool-core clusters.}
}
  {}
   \maketitle

\section{Introduction}\label{sec:intro}

Despite early claims that clusters of galaxies are straightforward systems to model, steadily improving observations as well as decade-long theoretical and computational efforts indicate that they are anything but.
There is, as of yet, no clear consensus on how the remarkable diversity of observed cluster core properties, ranging from {"strong cool core"  to "extreme non-cool core" (i.e. central cooling times ranging from one to two hundred million years up to several gigayears)} and everything in between, has emerged.  In the case of cool core groups and clusters, powerful jets from central supermassive black holes (SMBHs) have long been suspected of injecting the required energy into the intracluster medium (ICM) to compensate for radiative losses and maintain global stability (see for instance \citealp{rephaeli_energetic_1995,binney_evolving_1995,ciotti_cooling_2001,babul_physical_2002,mccarthy_towards_2008}), and while today there is broad consensus that this is indeed what is happening (for a review, see \citealp{mcnamara_heating_2007,mcnamara_mechanical_2012,fabian_observational_2012,soker_jet_2016}), there are a number of critical details associated with this jet-heating picture (commonly referred to as \emph{radio-mode} AGN feedback) that have yet to be properly understood; of these, two particularly stand out.  

The first concerns the origin of the gas whose accretion onto the SMBH powers the jet:  Is it due to hot/Bondi accretion or a drizzle of cold clouds condensing out of the ambient gas in the cluster cores and free-falling onto the central AGN?  
For a detailed discussion, we refer the reader to \citet{prasad_cool_2015,prasad_cool_2017} and references therein.  
Here, we simply summarise the current state of affairs by noting that that several different lines of observational evidence seem to collectively favour the ``cold rain'' model and theoretical studies indicate that the cold clouds are expected to naturally form in the presence of AGN-induced turbulence.
The latter calls attention to a broader set of related questions regarding the impact of AGN feedback on gas accretion processes; for example, does the AGN simply expel the gas around the SMBH, leading to the shutdown of AGN activity for a period of time, or does it  promote other mechanisms of accretion onto the AGN that also contribute to its self-regulation?
In a recent study, \citet{cielo_backflow_2017} found that jet-induced gas circulation (\emph{backflows}) can funnel as much as $1$~M$_\odot$/yr to the innermost parsecs (see also \citealp{antonuccio-delogu_feeding_2010}). 

The second issue concerns the coupling between the jets and the ICM:  how does  a central SMBH, powering apparently narrow bipolar outflows, successfully manage to heat the gas in the cluster cores in a near-isotropic fashion?  In this paper, we focus on this latter issue.

Establishing precisely how bipolar AGN jets interact with and heat the ICM in the cluster cores to prevent cooling catastrophes has proven to be an especially vexing problem. \arifI{The issue has been the subject of numerous studies dating back to the early 2000s (c.f. \citealp{Reynolds_2001,Reynolds_2002,Omma_Binney_Bryan_Slyz_2004,Omma_2004}).  In what was the first systematic attempt to address this problem using a series of high-resolution, three-dimensional hydrodynamic simulations, }
\cite{vernaleo_problems_2006}, who considered the standard model where the direction of the jets is fixed, found that such jets only managed to delay the onset of catastrophic cooling, not prevent it \arifII{(see also \citealp{oneill_intermittentjets_2010})}. The primary reason for the failure was that once the initial jet had drilled through the cluster core and excavated a low-density channel, all subsequent jets took advantage of this channel to flow freely out of the core, carrying their energy with them. The authors concluded that some additional complexity is required to ensure more effective heating of the cluster cores by AGN jets.

Detailed X-ray and radio observations of individual cool core groups and clusters (hereafter collectively referred to as ``cool core clusters'' or CCC) offer intriguing hints about how nature has addressed the ``isotropy'' problem in real systems.  Many of the CCCs  show evidence of multiple generations of active and relic jets, radio lobes, and X-ray cavities whose angular positions/directions in the sky are misaligned with respect to each other.  Since these by-products of AGN jets trace a nearly isotropic angular distribution about the cluster centre, \cite{babul_isotropic_2013} argue that the associated heating should do the same.   Moreover, since the observations indicate directional changes on time-scales ranging from a few to a few tens of millions of years --- which is typically shorter than the core cooling time --- the jets ought to be able to heat and maintain the core in at least a global equilibrium configuration.

\arifI{There are three distinct categories of models proposed to account for the observed misalignment of successive generations of  jet-lobe-cavity features:\newline
(1) The first invokes jets interacting with, and being deflected by, dense clouds and filaments in the ICM (c.f.~\citealp{DeflectedJets_1999,Mendoza_2001,Saxton_2005,Prasad_turbulence_clouds_2018}).  This scenario has not been explored much because historically the ICM was assumed to be largely homogeneous.  \arifIII{However, there may be cause to revisit this model.   Observations show that the central galaxies in cool core clusters are typically surrounded by extended filamentary warm-cool gas nebulae (\citealp{Hatch2007,Cavagnolo2008,Wilman2009,McDonald_Halpha_2010} --- see also \citealp{Heckman1989,Crawford1999}). Further support comes from theoretical and simulation studies, which strongly indicate that the ICM in this region ought to be replete with the accumulated detritus of cool gas drawn out of the central galaxies by AGN jets and bubbles (c.f. \citealp{Saxton2001,Bruggen2003,Revaz2008,pope_transport_2010,DuanGuo2018} and references therein), besides being locally thermally unstable and susceptible to  {\it in situ} cloud/filament formation (c.f. \citealp{
Cowie1980,Hattori1995,Heckman1989,Pizzolato_AGNClouds_2005,Pizzolato_AGNClouds_2010,NipotiBinney2004,MallerBullock2004,
McDonald_Halpha_2010,Hobbs2011,Sharma_TI_2012,gaspari_cca_2013,Li_TI_2014,Gaspari2017,Voit2017,prasad_cool_2015,prasad_cool_2017}, and references therein).}\newline 
(2) The second class of models invoke ``ICM weather'', that is,~wakes, bulk velocities, and turbulence on scales of a few kiloparsecs or larger, induced either by mergers or orbiting substructure (\citealp{soker_jetbending_2006,heinz_answer_2006,morsony_swimming_2010,mendygral_mhdjets_2012}). 
One potential issue with this class of model is the need for substantial velocity shear across the cluster core.  In a simulation study of 63 clusters by \cite{lauetal_perseusar_2017}, only a small fraction meet this bar.\newline  
(3) The third class invokes occasional changes in the orientation of the spin axis of the SMBHs that are powering the jets (and hence, the re-orientation of the jet axis).  This can occur as a result of precession and slewing (or tilting) of the black hole spin axis, particularly in combination with intermittent jet activity, and spin flips (c.f. \citealp{Merritt_SpinFlip_2002,Pizzolato_BBH_2005,Gitti_PrecessingJet_2006,Dunn_BHPrecess_2006,LodatoPringle_2006,Campanelli_SpinFlip_2007,Sternberg_Soker_FatBubbles_2008,Kesden_SpinFlip_2010,Precess_Jets_2010,Merritt_BHPrecess_2012,babul_isotropic_2013,Gerosa_BHPrecession_2015,Franchini_BHprecess_2016,Nawaz_JetPrecess_2016}).
\arifII{From a macroscopic, cluster core scale perspective, all of these are similar in that they give rise to jets whose orientation changes stochastically.}  Interestingly, a recent multi-wavelength study \citep{osullivan_giant_2012} of the core of z=0.442 cool-core galaxy cluster CL 09104+4109, and the Type II quasi-stellar object (QSO)
\LEt{Please spell out all acronyms the first time they appear in the paper, followed by the abbreviation in parentheses, both in the abstract and again in the main text. After that, please only use the abbreviation. See A and A language guide Section 5.2.4 www.aanda.org/language-editing} B0910+410 at its centre, offers tantalising support for this scenario.  B0910+410 is one of only two $z<0.5$ QSOs at the centre of a galaxy cluster and it seems to have switched from being a radio AGN to a QSO about 200 ${\rm Myrs}$ ago in response to a significant inflow of gas, and appears to be transitioning back to a radio AGN, with the bud of a new jet clearly misaligned with respect to the old large-scale {\it relic} jet.}

\arifI{In this paper, \arifII{the first of a series,} we use numerical simulations to investigate the latter class of models within the framework of a cool-core galaxy cluster with an initial central cooling time of $150$~ Myr.  We attempt to capture the basic feature of the different models within this category via stochastically re-orienting jets.  We investigate explicitly the extent to which such jets couple to the cooling ICM in the cluster core, \arifII{the efficacy of these models in affecting isotropic heating in the cluster cores,} and more broadly, the impact of such jets on the thermal and dynamical evolution of the hot diffuse ICM.  We explore three different prescriptions for jet re-orientation (also running a case without jets and a jetted but non-reorienting one, for reference), and compare the predictions for the properties of the bubbles and the stability of the cool core in each case.}

In Sect. \ref{sec:runs}, we describe the model we use for our re-orienting jets in a CCC, and the set-up of our simulations.
Sections \ref{sec:method}, \ref{sec:description}, \arifII{and \ref{sec:zoom}} are devoted to a detailed but qualitative description of the physics and appearance of the cavities and all the visible structures, obtained by comparing realistic X-ray images from the simulations with the physical state of the gas. We also discuss the role of projection effects.
Section \ref{sec:pdV} contains a quantitative analysis of the bubbles' properties and the heating that re-orienting jets are able to provide to the core, while Sect. \ref{sec:stability} relates these properties to observable large-scale inflows and outflows, to the temperature and stability of the cool core, and to the energy balance of X-ray gas.
In Sect. \ref{sec:discussion}, we relate our \sI{findings to those of} \arifII{previous simulation studies and discuss how our results would change \sI{by varying} the inactivity duration during a jet cycle}. In Sect. \ref{sec:conclusion}, we draw our conclusions on how re-orienting jets impact the shape of the X-ray cavities, the ICM as a whole, and the halo core. A following paper featuring the same simulations will be dedicated to an analysis of energy generation and transport, differentiating the effects of the different physical mechanisms (radiative cooling, shock-heating, advective or convective transport, mixing, turbulent dissipation, etc.), and how these determine the halos' gaseous profiles. In the following, we refer to this work as paper II.

\section{Models and numerical implementations}\label{sec:runs}

We test the reorienting jets model using a total of five numerical simulations. These simulations were run using the hydrodynamical, \emph{adaptive mesh refinement} (AMR) code FLASH v4.2 \citep{fryxell_flash:_2000}, adopting a modified setup described in \cite{cielo_backflow_2017}. In our computational setup, FLASH solves the non-relativistic Euler equations for an ideal gas, with specific heat ratio $\gamma = 5/3$, \arifII{initially placed in hydrostatic equilibrium within a gravitational potential well.    The gravity acting on the gas is that due to the gas itself as well as a static, spherically symmetric, dark matter halo (see below for details).} 

\arifII{The metallicity of the gas is set to [Fe/H]=-0.1 throughout and over the course of the simulation, the gas is subject to radiative cooling following the same prescriptions as used by \citet{cielo_backflow_2017}.  Specifically, we use the cooling function of \citet{sutherland_cooling_1993}, extended to higher  plasma temperatures (i.e. $\sim 10^{10}$ K) as described in Appendix B of
\citet{antonuccio-delogu_active_2008} to allow for a proper treatment of gas in the jet beams and the cavities.}

\arifII{We allow the \LEt{Please spell out all acronyms the first time they
appear in the paper, followed by the abbreviation in parentheses, both in
the abstract and again in the main text. After that, please only use the
abbreviation. See A and A language guide Section 5.2.4 www.aanda.org/language-editing}AMR in FLASH to refine up to level 10 (i.e. it refines at most ten times), if density and temperature gradients require so\footnote{The refinement criterion used is the same as in \citet{cielo_backflow_2017}, i.e. FLASH's default refinement strategy based on L{\"o}hner's error estimator \citep[see FLASH user manual, or][]{lohner_adaptive_1987}. We apply the criterion to both density and temperature, and set this parameter to $0.8$ for refinement and $0.6$ for de-refinement.}. At each refinement operation, a \emph{block} of interest is split in two along every spatial dimension. In addition, all the blocks are further divided into eight computational \emph{cells} along each dimension, giving a resolution element of $4\;{\rm Mpc}/(8 \times 2^{10})\simeq 488$~pc. Refinement can be triggered, up to maximum level, anywhere in the simulation box, without geometrical restrictions; for instance, in refining we do not privilege the central region over the jet-inflated cavities.}

\subsection{Initial conditions}
All simulations feature the same initial conditions meant to reproduce a CCC comparable in mass to the Virgo Cluster. 
We assume a flat background cosmology corresponding to $\Omega_m h^2=0.1574$, $\Omega_b h^2=0.0224$ and $H_0=0.7$ (\citealp{komatsu_seven-year_2011}), and use a static, spherically symmetric gravitational potential for an NFW halo ($\mathrm{M}_{200} \simeq 4.2\times10^{14}$\msun, $r_{200}\simeq1.7$~Mpc) to define our cluster. Guided by the mass-concentration relation by \cite{newman_density_2013}, which takes into account the presence of a central BCG (see their Sect. 10.1), and the recent analysis of the Virgo cluster observations by \citet{simionescu2017virgo}, we set our cluster concentration parameter to $c_{200}=10$.
The halo choice defines our simulation box.  In order to encompass the halo $r_{200}$, we use a cubic box of $4$~Mpc side. 

Having defined our dark matter halo, we next add to it a spherically symmetric hot gas component whose radial profile is subject to the following three constraints.

\begin{description}
  \item[] {\bf Initial entropy profile: } \arifII{For the starting entropy profile of the ICM in our simulations, we adopt the functional form $\ln(S(r)) = \ln(S_0) + \alpha \ln(r/r_c)$, where $S(r)\equiv k_BT(r)/n_e(r)^{2/3}$, $k_B$ is the Boltzmann constant, $T$ is the gas temperature, $n_e$ is the gas electron density, calculated by assuming fully ionized plasma with the given metallicity, and the power-law index   
$\alpha=1.1$ when $r$ is larger than the core radius $r_c$, and $\alpha=0$ otherwise.  This simple functional form has previously been  used to describe the observed diversity of entropy profiles across the cool-core/non-cool core spectrum \citep{cavagnolo_intracluster_2009} as well as starting configurations in theoretical and numerical studies \citep{babul_physical_2002,mccarthy_towards_2008,prasad_cool_2015}.  We choose a core radius of  $r_c=12$~kpc and a core entropy $S_0=12$~keV~cm$^2$, which results in a core with a cooling time of $150$~Myrs and qualifies our simulated cluster as a cool-core system.
We appreciate that recent studies show that groups and clusters with short central cooling times do not have isentropic cores, and instead exhibit an $r^{2/3}$ profile \citep{Panagoulia_CoreEntropyProfile_2014,CLoGS_GroupsEntropyProfile_2017,Babyk_entropyprofile_2018}.  Since we do not couple the jet activity to the state of the ICM (c.f. \S2.2), the detailed structure of the inner entropy profile has no bearing on our primary objective, which is to investigate the efficacy of the re-orientating jets at affecting near-isotropic heating in the core region.  For this, we only require that the core cooling time is shorter than the simulation run time.}
   
  \item[] {\bf Initial hydrostatic equilibrium:} We require the gas to be in hydrostatic equilibrium (HSE) within the dark matter potential. The HSE equation with the chosen entropy profile is not analytically integrable, so we numerically integrate it separately and import the tabulated profile into FLASH (the spatial sampling of the integration is chosen equal to the grid used in the simulations).
  \item[] {\bf Profile normalization}: In order to get the right hot-gas-to-dark-matter ratio for the given halo mass, we normalize the profile so that the ratio of hot gas to dark matter mass at the radius $r_{500}\simeq1$~Mpc is set to $60\%$ of the cosmic value, in agreement with the observational results shown in \cite{liang_growth_2016} for Virgo-mass systems.
\end{description}
The resulting gas temperature is between one and a few kiloelectron volts throughout.

\subsection{Reorienting jets: parameters and implementation}\label{sub:tiltSimulation}

Our implementation of the jet source terms is essentially the same as in \citet{cielo_backflow_2017}.  There, the bipolar jets were introduced as source terms within a rectangular prism consisting of eight central cells (four cells per beam) whose long axis was aligned along one of the axes of the simulation grid.  The only difference here is that we allow the inclination of the jet axis with respect to the simulation grid to vary. This, in turn, means that the number of injection cells also varies with the jets' inclination. We parametrize the jets' orientation using standard spherical coordinates $\theta$ (angle between the z axis of the grid and the jet axis) and $\varphi$ (angle of the positive direction of the x axis with the jet axis projection on the z=0 plane). When jets are injected at an angle, the injection cells and the momentum direction are changed to follow that orientation.

We present four simulation runs with jets and a control run with no jets. \arifII{In the jetted runs, we do not couple the triggering of the jets or their power to the state of the ICM.}  All four jetted runs are identical except for the prescription for their re-orientation angles, meaning that all jets have the same power, density, internal energy and injection base radius. 

Following  \cite{cielo_3d_2014}, the jets' density $\rho_{jet}$ is set equal to $1/100$ of the central halo gas density (see also \citealp{perucho_deceleration_2014,guo_importance_2016}).  Further, we fix the jet kinetic power to $P_{jet}=10^{45}$~erg/s, in agreement with measures of mechanical luminosities from X-ray cavities in galaxy clusters (e.g. \citealp{hlavacek-larrondo_extreme_2012}) and approximately equal to the total radiative losses of the halo gas. 

Besides their kinetic power $P_{jet}$, the jets also have internal energy, the flux of which, $U_{jet}$, can be simply computed from the jet \sI{Power $P_{jet}$ and the jet's \emph{internal Mach number} $\mathcal{M}_{jet} := v_{jet}/\sqrt{\gamma {p_{jet}}/{\rho_{jet}}}$, which we set to $3$:
\begin{equation}
U_{jet} = \frac{P_{jet}}{\mathcal{M}_{jet}^2}\frac{1}{\gamma \left( \gamma-1\right)} = \frac{2 P_{jet}}{9} \frac{9}{10} = 0.2 P_{jet} \label{eq:Ujet}.
\end{equation}
}
The total energy flux is therefore a constant $1.2\times10^{45}$~erg/s whenever the jets are on, and zero otherwise. All jets follow the same on/off schedule: each jet event lasts $40$~Myr, followed by a $2$~Myr quiescent period, during which the jet source terms are switched off.  After that, another jet event starts, generally along a different direction.  Overall, this means that feedback is active with constant power for $\sim95\%$ of the time.  Our choice of 40/2 Myr for the timing of the jet on/off cycle is guided by  the typical timescale between misaligned jet events observed in galaxy clusters, as catalogued by \cite{babul_isotropic_2013}, as well as the characteristic jet alignment timescale in their preferred physical model.

In this work involving constant-power periodic jets, we mainly explore the parameter space of reorientation angles, for which we adopt the following prescription: the jet polar angles $(\theta,\,\varphi)$ are chosen at random (from a spherical distribution) in a given interval of angular distance with respect to the previous jet axis; we use this interval to label our simulation runs. For example, in run 2030, the axis of any jet will form an angle (chosen at random) between $20$ and $30$ degrees relative to the previous jet axis. The runs we present are 0000 (i.e. jets along a fixed direction), 0030, 2030, and 0090 (given the bipolar nature of the jets, 0090 means the new direction is chosen totally at random on the sphere, with no constraints).

Table \ref{tab:angles} lists \arifIII{all orientation angles (polar, azimuthal) or ($\theta$, $\varphi$)  of each jet (in degrees, rounded to the nearest integer) for all simulation runs. Random numbers are drawn from a spherical distribution; then each direction is chosen so that its angular distance with respect to the previous jet, expressed again in degrees, lies within the interval that labels the run.}

\begin{table}
\caption{Orientation angles ($\theta$, $\varphi$) of all jets in each run.}
\label{tab:angles}      
\centering                                      
\begin{tabular}{rrcccc}          
\hline\hline                        
$\mathrm{Jet}$ &$t_{on}$ & 0000 & 0030 & 2030 & 0090 \\
\# &$\mathrm{Myr}$ & $\mathrm{deg}$& $\mathrm{deg}$& $\mathrm{deg}$& $\mathrm{deg}$ \\
\hline                                   
    1 &0   &(0,0) &(0,0)    &(0,0)    &(0,0)    \\
         2 &42  &(0,0) &(12,0)   &(25,0)   &(75,0)   \\
     3 &84      &(0,0) &(15,173) &(44,167) &(36,96)  \\
     4 &126 &(0,0) &(28,150) &(69,160) &(18,160) \\
     5 &168 &(0,0) &(41,161) &(95,136) &(72,177) \\
     6 &210 &(0,0) &(65,138) &(73,137) &(82,16)  \\
     7 &252 & -    & -       & -       &(75,173) \\
\hline                                             
\end{tabular}
\end{table}

We note that the first jet is always along the z-axis, while the second lies always in the $y=0$ plane for visualization simplicity; as the halo's initial conditions are spherically symmetric, this is just a choice of reference frame.

The run duration for the jetted simulations ranges from 227 to 277 Myrs; all simulations terminated during the sixth jet event, except run 0090, which reached the seventh jet event.  The few hundred million years duration of the simulations covers comfortably the initial central cooling time of the halo of $150$~Myr, so that in the no-jet run a large cooling flow develops. The number of jet events we covered is in principle not free from statistical under-sampling; however, a visual inspection of the runs reveals that the intended solid angle coverage is achieved in all cases. For instance, in run 0090, cavities never substantially overlap, so that jets are indeed affecting a portion of solid angle that is as large as possible; on the contrary, in run 0030, the jets quite often end up in the trail of the previous cavity, as we see in Sect. \ref{sec:description}.

\section{Cavities: physical background and method}\label{sec:method}
Our refinement criterion keeps the jet/cavity system maximally refined at all times, therefore providing detailed insight on the physics of the bubbles, including velocity and turbulent structure, rise and expansion in the external gas, and the implications for the energetics of the cool-core halo.

We are also able to pair this physical view with synthetic observations, via the production of realistic X-ray emissivity maps (see Sect. \ref{sub:method} for details), in order to provide direct comparison of the individual features from X-ray observations of galaxy clusters. 
The morphology of real X-ray cavities is often more complex than a collection of bubble pairs (e.g. \citealp{zhuravleva_nature_2016}), as several other physical processes  are at work. Some are  due to the bubbles themselves: shocks, both weak and strong, \citep{nusser_suppressing_2006}, as well as the 
generation and dissipation of pressure waves (\citealp{stemberg_soundwave_2009,fabian_transport_2017} and references therein) during the inflation of the bubbles; wakes, ripples and ICM motions excited by buoyantly rising bubbles \citep{churazov_calorimeter_2002,nusser_suppressing_2006}, lifting of and subsequent mixing with low-entropy gas \citep{bruggen_bubbles_2003,pope_transport_2010}, and so on. Other processes are due to the cluster environment: shocks, cold fronts, streams and other transient structures associated with mergers~(\citealp{poole_mergers_2006} and references therein), tails of diffuse ionized gas ram-pressure stripped from cluster/group galaxies (e.g.~\citealp{boselli_ionized_2017}), and so on.
Many of these processes can be isolated and studied simply by processing the X-ray images (see \citealp{churazov_arithmetic_2016}), while simulations can provide insights about the origins of the various features.

\subsection{Structure generated by a single jet}\label{sub:onejet}

Several numerical studies (e.g. \citealp{vernaleo_problems_2006,sutherland_interactions_2007,cielo_3d_2014}) have investigated the evolution of AGN jets propagating through an unperturbed ICM.   The evolution follows a characteristic trajectory that also provides a fitting description of the first jet event in our simulations.  Below, we briefly summarise the main features of this trajectory as it unfolds in our runs.

The initial interaction between a jet and the ambient gas creates a $\sim10^{10}$ K hot spot (HS; usually no wider than $1$ or $2$~kpc) and a bow-shock propagating for several tens of kiloparsecs. 
The two bow-shocks enclose an ellipsoidal \emph{cocoon}, filled with sparse, hot, and turbulent gas. Internally, the expanding cocoon supports large-scale gas circulation that results in \emph{backflows}, while globally 
it behaves like an almost uniform overpressurized bubble.

Near the HS, the shocked gas collects in two very hot cavities, which in a few million years evolve into lobe-like structures, typical of classical radio galaxies. The gas in the ``lobes'' is  denser and hotter than in the rest of the cocoon; \cite{cielo_3d_2014} name this the \emph{lobe phase}.  The jets then switch off: the bow-shocks lose their drive and 
slow down, first becoming transonic and eventually subsonic,
while the lobe gas detaches from the centre, forming hot low-density bubbles. 


The rising bubbles retain their inner velocity structure, a \emph{vortex-ring-like} bubble-wide circulation, as expected for light, supersonic jets (see \citealp{guo_shape_2015}).  The motion of the bubble-ICM boundary due to vortices inside the bubbles and the backflow of the ICM around the bubble excite sound waves \citep{stemberg_soundwave_2009}.  As the bubbles ascend and move into regions where the ambient gas pressure is lower, they expand and cool. Adiabatic cooling dominates over radiative cooling because of the bubble's low density. 

The second and subsequent generations of jets that follow propagate through an already perturbed ICM, so their evolution is strongly impacted by encounters with structures generated by earlier jets.
For example, AGN jets possess high velocities but low inertia: dense older bow-shock fronts can act like walls, deflecting the jet beams and exciting oblique shocks  or ``ripples'' in dense regions of the surrounding gas; channels and cavities carved by previous jets act as low-resistance conduits for subsequent jet flows.

\subsection{Method: physical properties of the gas 
\LEt{Please spell out all acronyms the first time they
appear in the paper, followed by the abbreviation in parentheses, both in the abstract and again in the main text. After that, please only use the abbreviation. See A and A language guide Section 5.2.4 www.aanda.org/language-editing} versus X-ray maps}\label{sub:method}
Figures \ref{fig:0000} to \ref{fig:0090} show three different panels for a single snapshot of each jetted run, presented from the least to the most isotropic jet distribution, that is, 0000, 0030, 2030 and 0090.

By comparing those figures, we observe the different predictions for location, visual aspect, and physical state of the hot bubbles in the cluster's gaseous halo. 
All images still present a very high degree of central symmetry; this is due to the absence of substructure or asymmetry in our initial halo, or the absence of a central galaxy (since a central clumpy ISM may induce asymmetry in radio jets, as shown by \citealp{gaibler_asymmetries_2010}). This is not necessarily true in real CCC cavities, nonetheless most of the complex features we observe in our figures can be directly compared with the observations and are indicative of the various physical jet-jet and jet-ICM interactions taking place in the central few hundred kiloparsecs. On scales of a few megaparces, cosmological accretion as well as the interaction of the ICM with infalling substructure can give rise to shocks and pressure waves (see, e.g. \citealp{poole_mergers_2006,storm_mergingcluster_2015}) but these can be easily distinguished from features originated by AGN feedback.

In the following, we describe the content of each panel in Figs. \ref{fig:0000}-\ref{fig:0090}, from top to bottom.

\paragraph{\bf Three-dimensional rendering of the gas temperature}obtained with a ray-casting technique. \arifII{The line of sight is along the X-axis.} The X-ray gas background has been made transparent (as indicated by the opacity annotation next to the colour key in each plot) in order to highlight jets and cavities. From coldest to hottest, we can recognise: 
\begin{itemize}
\item{\emph{the halo's cool core}}, extending up to a few tens of kiloparsec and presenting the lowest temperatures in the simulation (about and below $10^7$~K); 
\item{\emph{the bow-shocks}}, around $10^8$~K (although they cool relatively rapidly to the background gas temperature) are generally more visible only around the most recent jets; 
\item{\emph{the jet-inflated bubbles}}, ranging from a few $\times\;10^7$ to $\sim 2\times10^8$~K  (the younger, the hotter);
\item{\emph{the latest jet beams/Hot Spots}}, as hot as a few $\times\;10^{10}$~K, depending on the jet age.
\end{itemize}

The temperature of the cavity gas shown in the three-dimensional (3D) renderings is a much more  reliable proxy of their age than their volume or their projected distance from the cluster centre.   In the presence of multiple bubbles, the combination of projection effects and unknown relative orientations of the jets that gave rise to them makes age estimates based on volume and distance highly uncertain.

\arif{Admittedly, there are instances where young jet material ends up in old bubbles.  However, even in this case, it is possible to discern plasma of different temperatures within the cavity and consequently, a qualitative relation between the age of the plasma and its temperature (or energy, should the emission be non-thermal, as in the case of pure synchrotron) holds. Radio or hard X-ray observations (see Sect. \ref{sec:zoom}) ought to be able to effectively constrain the re-orientation history.} 

\paragraph{\bf Synthetic X-ray maps.}
The middle panel of each figure presents synthetic soft-X-ray observations, in which jets and bubbles appear as voids, providing a view complementary to the temperature. In order to make these images as realistic as possible, we generate them by processing our simulation output with the pyXSYM software (based on the work by \citealp{biffi_investigating_2013}). The software computes thermal and line emission from the hot gas in the X-ray band, then it generates and propagates the corresponding individual photons through the simulation domain.  \arifII{The projection is along the X-axis.}

For these maps, we choose an X-ray band of $[0.5,\,7.0]$~keV, then set all the sources at redshift $z=0.02$ and collect the photons for $1$~Ms from a $6000$~cm$^2$ telescope area. These values are chosen to match the specifics of a realistic observation with the \emph{Chandra} telescope, except for the telescope area, 
which for our mock observations is about ten times larger than the one of the ACIS-I detector on \emph{Chandra}.
Typical background counts (about $0.76$~photons/$\rm{cm}^2$/second) are added, but turn out to comprise less than $1\%$ of the total signal. Galactic absorption is also included with a Tuebingen-Boulder model (see \citealp{wilms_absorption_2000}). The generated photons are then collected on the simulated detector; spectra and images can be obtained at this stage. Once we obtain the raw images, we apply a standard \emph{unsharp mask} filter, as is sometimes done in the literature, to emphasize structure 
\sIII{of a specific size; this filtering operation is not included in pyXSIM. The filter does not preserve the photon count in each spaxel; however, here we are mostly concerned with the visibility and appearance of the cavities rather than flux measurements. In these initial X-ray images, we do not simulate any specific detector response, but just collect all generated photons. As a by-product, the image retains some visual imprint of the original grid.}
All images have a rather bright core and a
consequently reduced contrast in the peripheral regions, due a to high central cooling luminosity (as in run 0000) or to the latest bow-shocks (as in 0090).
The youngest bubbles are distinguishable most of the time, although they can sometimes be concealed by older bubbles or outshone by a bright bow-shock.

\paragraph{\bf Pressure slices.} Finally, in the bottom panels of Figs. \ref{fig:0000} to \ref{fig:0090} we show central plane slices of the gas pressure (expressed in internal units in order to avoid numerical rounding errors; our unit corresponds to about $3.9\times10^{-15}$~Pascal). While the first two panels are projections along the $x$ direction, these slices are contained in the plane $x=0$, the viewing direction being $y$ this time, in order to provide a different point of view. Only the structure in the central plane (in which the first two jet beams lie) are visible. The gas pressure clearly shows the waves and shocks that leave imprints in the X-ray gas and allow us to track their origin\footnote{However additional work is required to distinguish weak shocks from pressure waves, as we will show in paper II.}.  The full time-evolution movies of the pressure slices, provided as additional material to this paper, are very instructive in this respect.

\paragraph{\bf Labels.} To facilitate discussion to follow, the various structures in the panels in each of the figures are annotated and labelled.  We use the letters ``B'', ``C'' and ``R'' to denote bow-shocks, cavities (both lobes and bubbles) and \emph{ripples}, respectively.  By ``ripples'', we are referring to those complexes of weak shock fronts appearing in some of our X-ray maps, mostly near the core and in the vicinity of the youngest bow-shocks.  All recognised features are numbered sequentially from oldest to youngest; features associated with the same jet event, based on our analysis of the time-evolution movie, are assigned the same number.  The numbers \emph{do not necessarily} refer to the jet numbers listed in Table \ref{tab:angles}, as we only label what is visible in the last panel, meaning that some jets may be skipped. Bilaterally symmetric structures are only labelled on one side.


\section{Results: physics of cavities and their appearance}\label{sec:description}

\subsection{Run 0000}\label{sub:0000}

\begin{figure}
  \centering
  \includegraphics[width=0.43\textwidth]{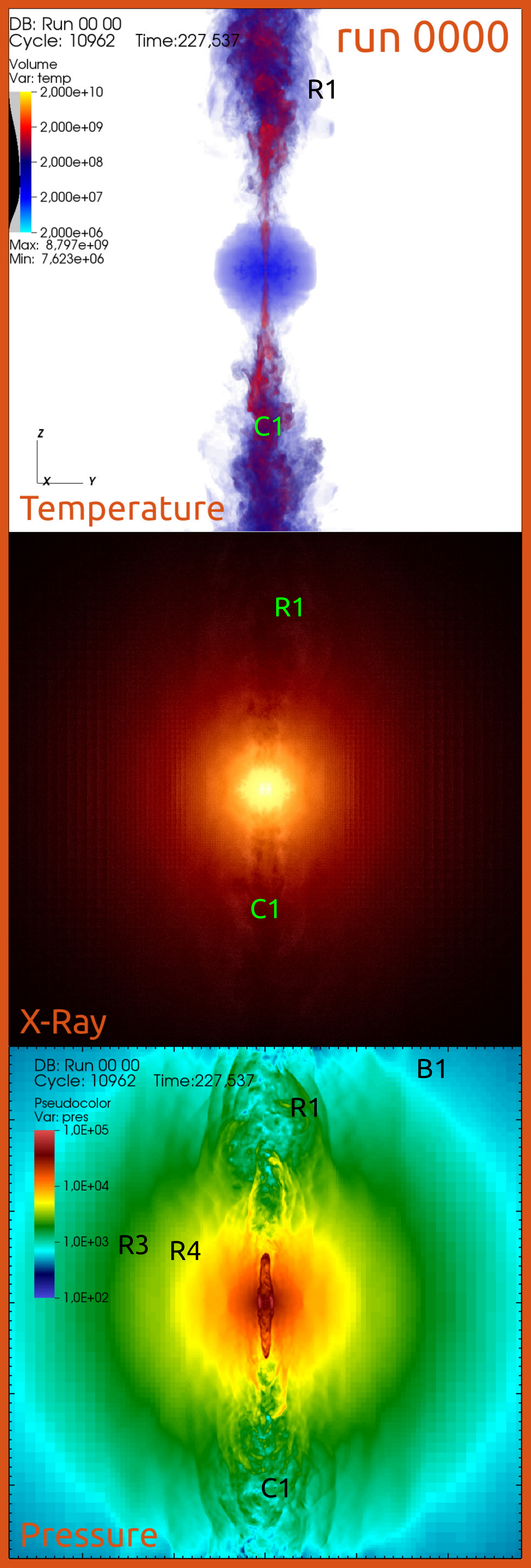}
\caption{Run 0000: 3D temperature, focusing on cool core and jet material, line of sight parallel to the x-axis (top); X-ray, projection along the x-axis (centre), pressure slice in the $x=0$ plane (bottom). Size: $400$~kpc, linear scale. Labels mark cavities (C), bow-shocks (B) and ripples (R), numbered from oldest to youngest. See text and associated movie.}      
  \label{fig:0000}
\end{figure}

Run 0000 is shown in Fig. \ref{fig:0000}. Here all jets keep inflating the same cavity (labelled as C1): the $2$~Myr interval between two successive jet events is \arif{sufficiently short that there is not enough time for the channel carved out by the first jet to collapse, and all following jets follow this path of low resistance towards C1.}  The latter keeps \arif{moving further away from the centre and grows larger but never completely detaches from the centre to form a proper bubble because the channel is repeatedly refreshed by a new jet beam.} These connections are visible in the X-ray as rather large jet ``chimneys'' around the jet beams.

Most of the energy of the jets reaches C1, yet in the chimneys, several weak shocks take place. These include both self-collimation shocks of the beam, and internal reflections of the latter on the chimney walls. These shocks leave ripples in the X-ray gas, such as the ones marked as R3 and R4. In other words: a fresh jet encounters a dense medium, the chimney walls, with almost-zero attack angle, so the new beam gets almost totally reflected. The ripples are the only energy transmitted to the outer medium\LEt{A and A insists footnotes be incorporated into the body text when discursive and where possible.
Please incorporate this footnote into the text}. These perturbations are the main visible difference between a continuous and a pulsating beam. They propagate sideways, but weaken and fade in time, so their presence can be associated with a specific recent jet.  \arif{Another ripple structure occurs close to the boundary of the cavity, interior to the B1 bow-shock.   These ripples are excited by repeated inflation of the cavity by successive generation of jets, starting with the first one.}

The most peculiar aspect of run 0000 is the very large size of the bubble and its corresponding large-scale ($\sim 200-300$ kpc) bow-shock, B1.  \arif{In many ways, the complex jet/giant cavity/bow-shock resembles those of galaxy clusters Hydra A\footnote{see 
\hangafter=1\hangindent=10pt 
\url{http://chandra.harvard.edu/photo/1999/0087/more/0087_comp_lg.jpg}}
~\citep{nulsen_hydraA_2005,wise_hydraA_2007}
and 
MS0735.6+7421\footnote{see \hangafter=1\hangindent=10pt 
\url{http://chandra.harvard.edu/photo/2006/ms0735/}}
~\citep{mcnamara_heatingnature_2005,mcnamara_ultramassiveAGN_2009}.}
We see from the temperature panel that run 0000 is the one in which we can observe jet material the farthest from the AGN (beyond the $200$~kpc region shown here), given the ease with which it propagates within the old cavity, dispersing all over its volume. In summary, this model predicts emission from plasmas of different ages from one very large bubble.

\subsection{Run 0030}\label{sub:0030}

 \begin{figure}
  \centering
  \includegraphics[width=0.43\textwidth]{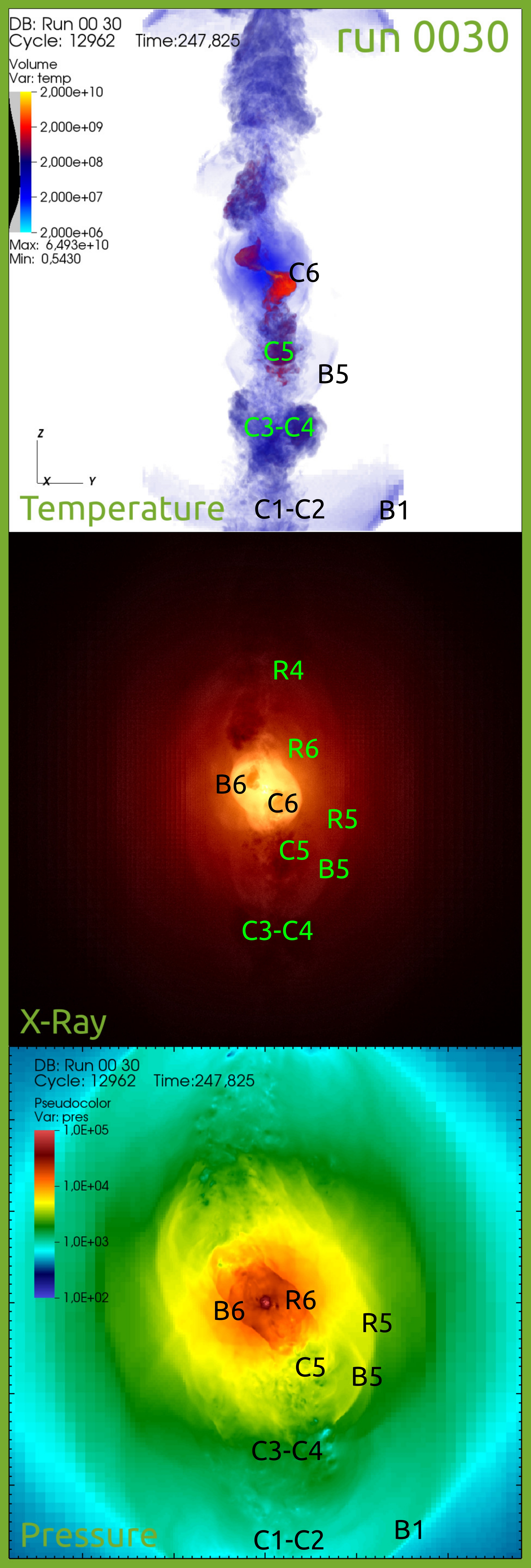}
\caption{As in Fig. \ref{fig:0000}, but for run 0030. We note how cavities are grouped together, as often two or more jets end up inflating the same bubble. See text and associated movie.}
  \label{fig:0030}
\end{figure}

In run 0030 (see Fig. \ref{fig:0030}), one can distinguish individual bubbles, but those are still part of a connected structure originating from the first jet axis, and appear as short thick branches of a main trunk. As a consequence, a significant fraction of the energy still flows through the first carved chimney. Proof of this is the main bow-shock, B1, almost as large as in run 0000.  

\arif{Of all the features, the brightest is the \emph{core + B6} complex;  the cavities C6 and C5 are also visible, and C5 shows an elongated morphology. 
Quite often, two successive jets end up within the same cavity; therefore we observe the frequent formation of composite bubbles, such as the features labelled as C1-C2 and C3-C4.} 

\arifII{Finally, we note that the conical structure of interconnected cavities that emerges in run 0030 is a generic feature of configurations where the jets generally change direction by small angles, regardless of the physical process responsible.  Composite cavity structure appears, for example, in the jet simulations of \citet{mendygral_mhdjets_2012}, in which the bulk flows in the ICM cause the jets to deflect by small angles, as well as in simulations of \citet{yang_how_2016}, in which intermittent jets precess about a fixed axis.   In both of these cases, the jets either end up intersecting and pushing into, or newly forming cavities end up breaking through and expanding into, pre-existing cavities excavated during earlier jet cycles.}

In terms of temperature, we can observe two populations of plasma of different ages, and only the most recent bubbles contain young jet plasma. \arif{The bleeding of one cavity into another  impacts their size evolution, which in turn will play havoc with the power estimations based on this measure.}  The external shape of older cavities in X-ray matches rather well the corresponding shapes in the temperature view, but provides no information about its internal structure. Overall, the X-ray map of this run resembles the numerous X-ray images of real galaxy groups and clusters that show a single pair of prominent cavities (c.f. 
Abell 2597\footnote{\url{http://chandra.harvard.edu/photo/2015/a2597/}};  \citealp{mcnamara_a2597_2001}). 

All cavities form secondary bow-shocks, but \arif{these are weaker and fainter} than the one associated with the very first jet, as some of the energy of the subsequent jets tends to flow along the first jet channel; of all the bow-shocks, only B6 (the youngest one) and, partially, B5 are visible in the X-ray image (they are both clearly visible in the pressure slice as well).

The youngest jet beam shows clear signs of deflection (in temperature): a bright hot spot and a ``plume'' shape, where the post-shock beam is deflected into the nearest C5 lobe. Plume-like structures like this one can \arif{induce} significant (asymmetric) backflows, as \arif{observed}, for instance, in \emph{X-shaped radio galaxies} (e.g. \citealp{roberts_abundance_2015}). The beam deflection is also \arif{responsible for exciting} ripple features (R5 and R6).

\subsection{Run 2030}\label{sub:2030}
 \begin{figure}
  \centering
  \includegraphics[width=0.43\textwidth]{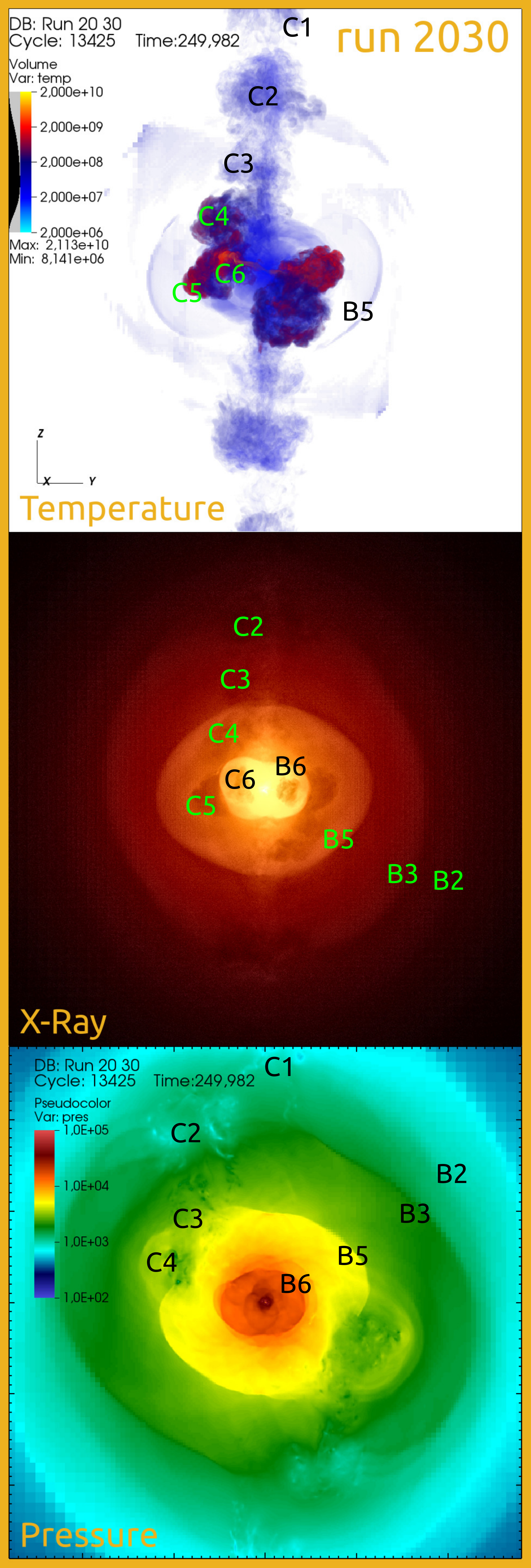}
\caption{As in Fig. \ref{fig:0000}, but for run 2030. This time bow-shocks and cavities are mostly distinct from one another, but subject to projection effects. See text and associated movie.}    \label{fig:2030}
\end{figure}
\arif{The first jet axis in run 2030 (Fig. \ref{fig:2030}) is, as in runs 0000 and 0030, clearly visible as a rising chimney of hot gas in both temperature and X-ray emission even after $250$~Myr; however, unlike the previous runs, the up-down channels are weaker and fading because they have not been reinforced by subsequent jet flows.}

\arif{The hot bubbles in run 2030 are also much more spatially spread out than in the previous two runs. The spatial distribution of cavities C1 to C4 in the X-ray and temperature maps may, at first glance, suggest that these cavities too are simply branches of a main trunk, but this is a projection effect, as confirmed by observing from a different direction (c.f. the pressure map). In fact, physically connected, composite cavities are no longer present, and all the bubbles can be individually identified and easily labelled, especially in the temperature view.}

The fact that the cavities are fully detached means that most of the energy of each jet is spent creating and inflating new bubbles closer to the halo centre, rather than inflating old, far-away cavities.
Therefore, jets and young bubbles continuously drive shocks near the core, keeping its internal energy higher.   The bubbles also spend more time within the innermost $100$~kpc (as C4, C5 and C6 in the shown snapshot) and are distributed over a rather large solid angle, reducing the space available for the formation of cooling flows. 
An intuitive explanation of why the bubbles are now detached from each other can be found by comparing this run with the 0000 case: the hot \emph{lobes} formed by fixed-axis jets define a (bi)conical region of jet influence (in agreement with previous numerical works), whose half-opening angle is roughly 15 degrees. In run 2030, the re-orientation angle is always forced to be larger than this value. Consequently, the new jets do not interact with the pre-existing channels and bubbles, and instead pierce the surrounding ISM/CGM along a new direction. The absence of ripple-like features further confirms this hypothesis, as ripples typically arise from interactions between the jet flows and the walls of previously formed channels and cavities. 

The three youngest cavities (C4 to C6 in Fig. \ref{fig:2030}) show a clearly higher plasma temperature, above $2\times10^8$~K, and are also the most visible in X-ray view (despite C4 being partially hidden by the brighter B5). Given that the cavities are physically distinct, one would expect that in this scenario the size and shape of the cavities can be straightforwardly extracted from the X-ray observations.  However, since most of the structure is concentrated in the central $100$~kpc, projection overlaps are likely (e.g. C4 and B5 in the X-ray map, or C5 and C6 in the X-ray and temperature maps).

The bow-shocks (or their slowed-down remnants) are detectable for longer than 100~Myr in both the X-ray and pressure maps as rather bright, sharp, clean fronts, except where they interact with pre-existing bubbles. One can distinguish almost all of them, from B2 to B6 (B4 is not visible in this particular snapshot due to its overlap with the B3 feature, but it is clear at earlier times).  The high brightness of the bow-shocks may however affect the measurements of the gas profile, and therefore have to be subtracted carefully from the image; but this is usually not an issue if multi-wavelength X-ray data are available (e.g. \citealp{churazov_arithmetic_2016}).

\arif{The images generated from run 2030 have a number of similarities with deep \emph{Chandra} X-ray observations of the cores of Perseus\footnote{\hangafter=1\hangindent=10pt
\url{http://chandra.harvard.edu/photo/2000/perseus/more.html}}$^,$\footnote{\url{http://chandra.si.edu/photo/2005/perseus/}}
by \citet{fabian_perseus_2000} and \citet{fabian_perseus_2011}, NGC 5813\footnote{\url{http://chandra.harvard.edu/photo/2015/ngc5813/}} by \citet{randall_ngc5813_2015}, and especially M87\footnote{\url{http://chandra.harvard.edu/photo/2006/m87/}}$^,$\footnote{\hangafter=1\hangindent=10pt 
\url{http://chandra.harvard.edu/photo/2008/m87/m87_xray.jpg}} by \citet{forman_M87_2005,forman_M87_2017}.
The deep X-ray image of the M87 reveals a series of loops and cavities that are thought to have been produced by a series of outbursts by a swivelling jet.}

\arif{
Finally, we point out that the pressure map for run 2030 shows good qualitative agreement with pressure disturbances seen in Perseus \citet{fabian_perseus_2011}.}

\subsection{Run 0090}\label{sub:0090}
\begin{figure}
  \centering
  \includegraphics[width=0.43\textwidth]{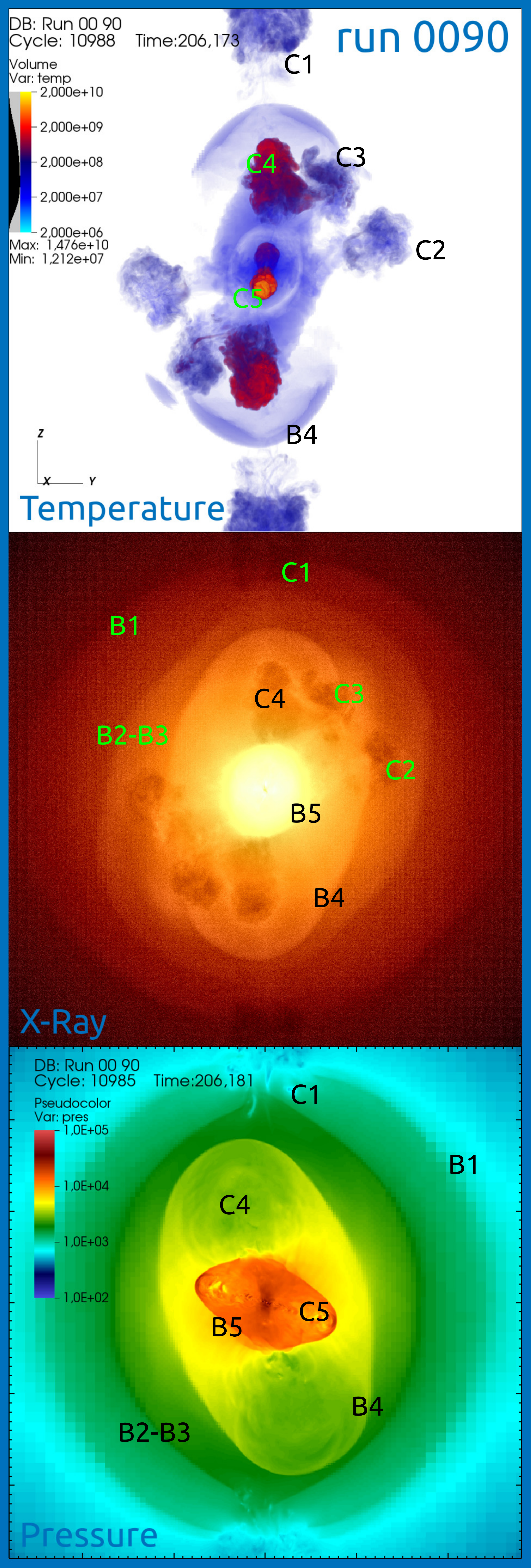}
\caption{As in Fig. \ref{fig:0000}, but for run 0090. This time, the bow-shocks are the brightest features, and one can see many cavities around the distance (projected) of $100~kpc$. See text and associated movie.}         
  \label{fig:0090}
\end{figure}
The bubbles in run 0090 (Fig. \ref{fig:0090}) are clearly detached and evolve almost completely independently; each jet creates its own bow-shock, and undergoes all evolutionary stages for single jet events described in Sect. \ref{sub:onejet}, virtually never interacting with pre-existing cavities. This is most clearly seen in the temperature and pressure views while in the X-ray, projection uncertainties are still significant, so some cavities appear joined together (e.g. the C3 and C4 features -- and possibly C2 -- can be mistaken for a single large bubble). The complex network of overlapping shocks and cavities makes it  very difficult to unambiguously establish the bubbles' ordering and energy unless additional information is available (e.g. temperature).  Moreover, there is a separate indication that the size of the cavities in the X-ray view may not be accurate: in the pressure slice, the cavity is much more extended than in the X-ray view.  This suggests that only the highest-contrast, central regions of the cavities stand out in the X-ray images (but recall that the viewing direction is different between X-ray and pressure).


\arif{Overall, the X-ray image is similar to the 2030 case, with well-defined bow-shock fronts as well as easily identifiable ghost cavities and nested cocoons, showing realistic positions and morphologies. The very bright centre in the X-ray map is due to the cocoon/bow shock of bubble being inflated by a jet aligned close to the line of sight, so that the hot spot points almost towards the observer.}

\arif{In X-ray, most of the bubbles (except the young C4 and C5, which still retain their elongated lobe shape) are roughly spherical. 
The inner structure of the cavity is visible only in temperature; a further indication that the shape of the cavities in the X-ray is not necessarily a good indicator of the physical nature of the cavity.  Even vortex-ring-like structures may appear almost spherical in X-ray images.}

The pressure panel provides a good view of the youngest jet beam (within C5), with visible individual recollimation shocks and terminal hot spots. This shows that shocks and jets are still effectively moving X-ray gas away from the central $100$~kpc, and with a solid angle coverage of almost $4\pi$.

\arifII{The large angle misalignment between the young jet beam and C4 cavities seen in the pressure panel closely resembles the jet and cavity structure seen in galaxy cluster RBS 797\footnote{\url{http://www.evlbi.org/gallery/RBS797.png}}.
\citet{Gitti_RBS797_2006} and \citet{Doria_RBS797_2012} interpret the combination of radio and X-ray data as suggesting that over the course of the three identified outbursts, the jet axis appears to have changed direction by $\sim 90^\circ$ between outbursts.}

\section{Zoom-in X-ray and projection effects}\label{sec:zoom}

\begin{figure*}
  \sidecaption
  \includegraphics[width=12cm]{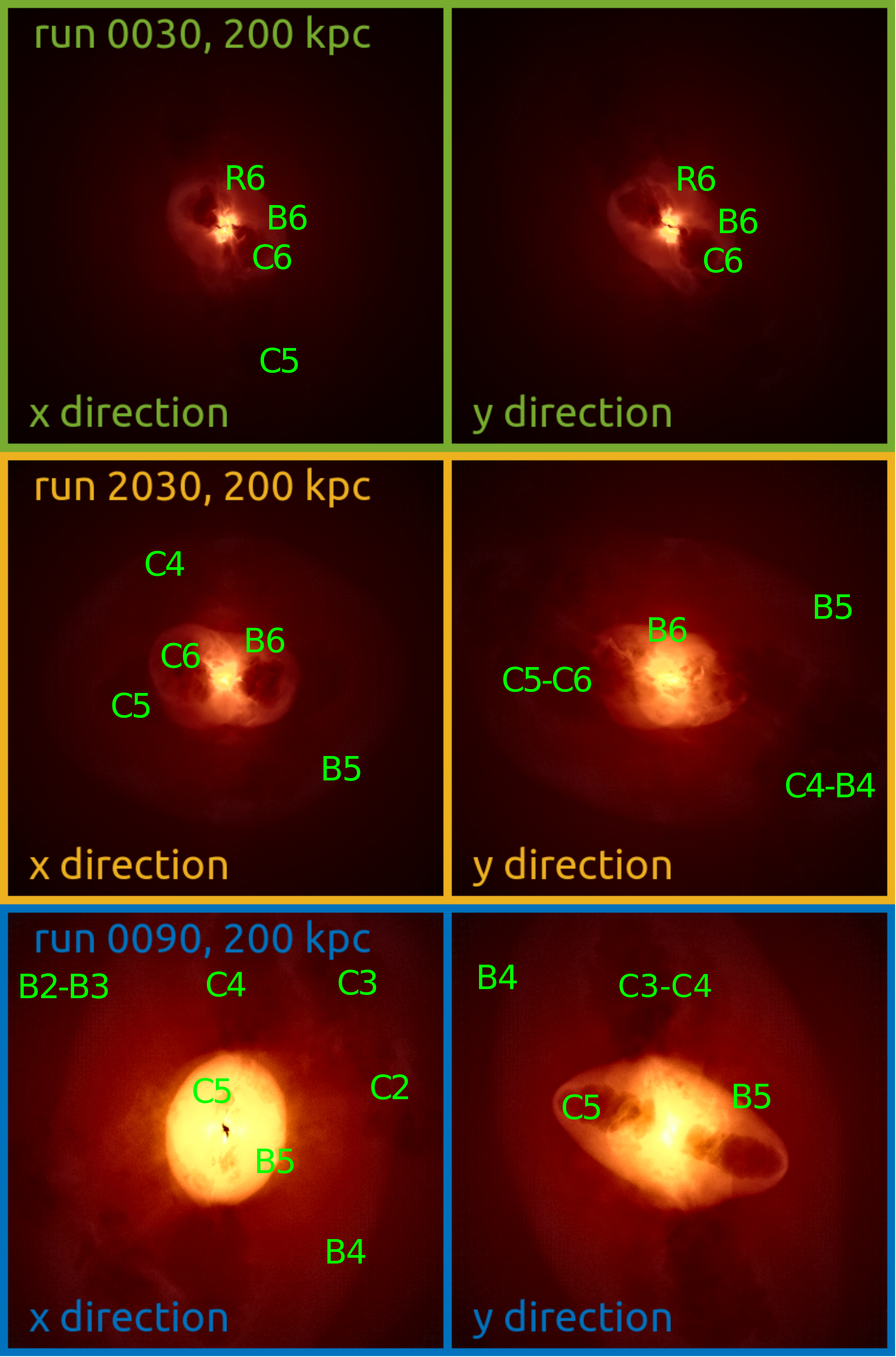}
  \caption{Synthetic X-ray observations, similar to the ones appearing in Figs. \ref{fig:0030} (run 0030) to \ref{fig:0090} (run 0090), but zooming into the innermost 200 kpc and simulating the ACIS-I detector. Projections  are along the $x$ (left column) and $y$ axes (right column). The labels match the ones in Figs. \ref{fig:0030} to \ref{fig:0090}, for direct comparison. We note the increased details at the cavity boundary and how the different perspective changes the apparent volume of the cavities.} \label{fig:xrayzoom}
\end{figure*}

We now present some more detailed X-ray views of the cluster core, 
which also exemplify the projection effect-related uncertainties in that crowded region. In Fig. \ref{fig:xrayzoom}, we show two different projections side by side, from the $x$ (left) and $y$ (right) direction, for each simulation run except the trivial case of 0000, where very little structure is present in the inner $200$~kpc.  The pictures are identical in all other respects.
Labels are put next to each feature, to facilitate direct comparison with Figs. \ref{fig:0000} to \ref{fig:0090}.

Each panel shows a zoom-in  X-ray image of the central $200$~kpc. This time, we use pyXSIM to simulate the ACIS-I detector (response, point spread function, field of view) on board the \emph{Chandra} X-ray telescope. As a side-effect of this operation, the photon counts are now overall lower, but the imprint of the simulation grid pattern is lost. The zoom-in also reveals some more details on the core and the innermost cavity structure; for example, in all images, the B5 and B6 features show structure in their shock fronts.  \arifII{Most of the earlier generation of cavities are now mostly invisible.}

In run 0030, the differences between $x$ and $y$ are minimal  (e.g. C5 and C6 are roughly in the same places), as the latest two jets have similar inclination with respect to the x and y axis. This is expected, as moderate re-orientation always produces structure aligned with the $z$ axis, and both views have the same orientation with respect to that.

In the 2030 and 0090 cases, there is no clear axisymmetry, so in some projections, smaller bubbles blend into the already dark background of the older ones.  In run 2030, for example, C5 and C6 appear distinct in the $x$ projection (separated by B6), but largely overlap in the $y$ projection. B6 is still visible as it is a young and bright feature, but it \arifII{is difficult to straightforwardly associate this feature with its corresponding bubble.}
\arif{If the overlap is only partial, compound cavities may appear as single, irregular cavities (e.g. C5 and C6 in run 2030/$y$ projection, and C3 and C4 in run 0090/$y$ projection).}

Young bubbles can also be outshone by the high brightness of their own bow-shock, if the line of sight is close enough to that jet's axis (as in run 0090, where B5 almost completely hides C5 in the $x$ projection). In the latter case, only a tiny portion of the cavity can be identified, so its volume and shape can not be used to infer jet properties, especially with a lower-quality image such as the one in Fig. \ref{fig:0090}.

\arifII{We also highlight the relationship between the degree of jet re-orientation and the brightness of the core: the greater the re-orientation, the brighter the core.  The core brightness is an indicator of shocks and hot, shocked gas. As we have noted previously, moderate-to-strongly re-orientating jets tend to inflate cavities that are more localised to the cluster core
(the labels in the panels for runs 2030 and 0090 reach lower numbers), and hence, this is also where the shocks associated with the inflating cavities appear.   The net result is more efficient heating of the core gas, which has deep implications for the halo stability, as we show in Sect. \ref{sec:stability}. }


\begin{figure*} 
  \centering
  \includegraphics[width=17cm]{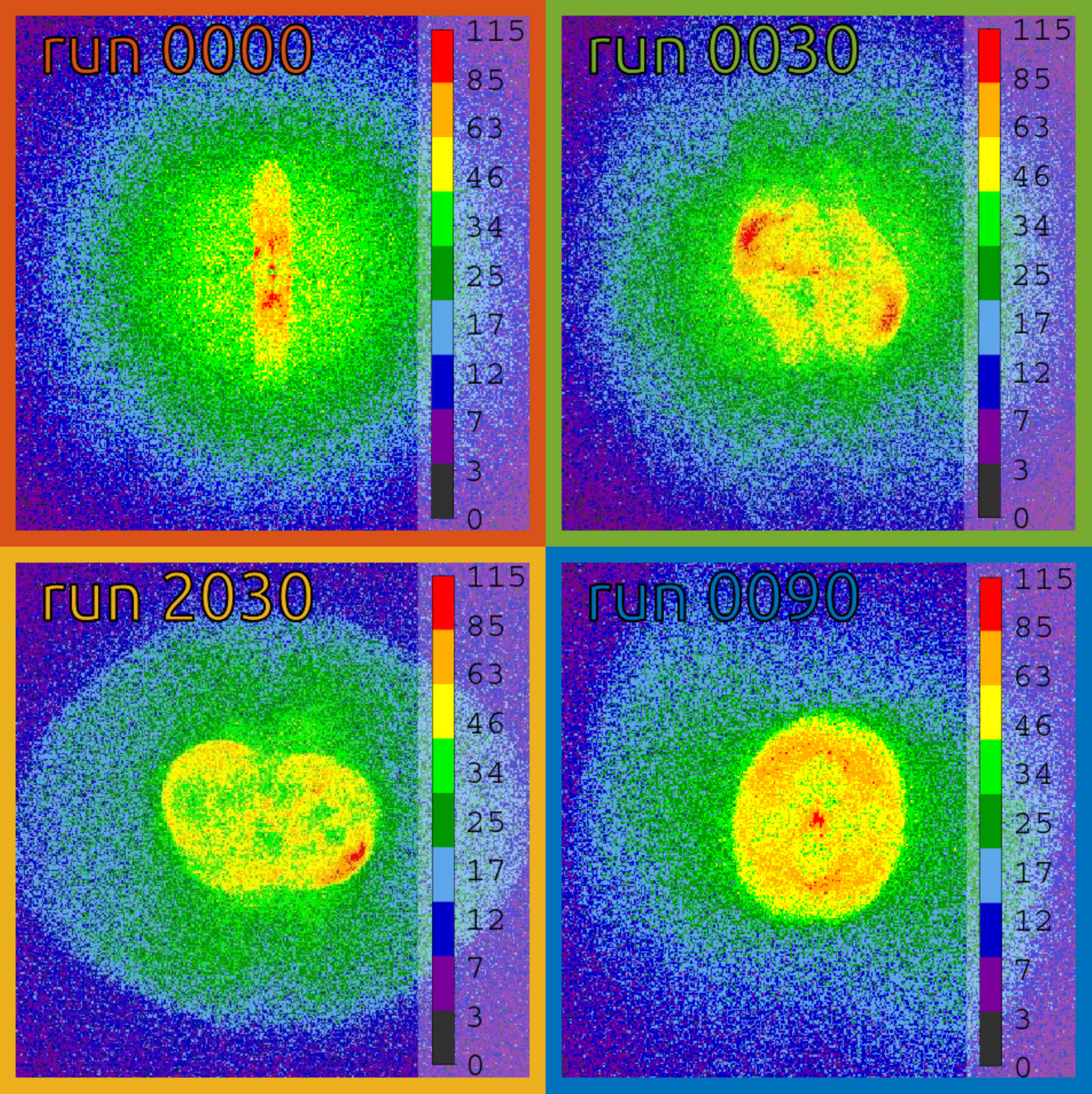}
  \caption{Synthetic hard X-ray observations, in the band $[10-30]$~keV, collecting all generated photons.  Projection is along the $x$-axis, as in Figs. \ref{fig:0000} - \ref{fig:0090}, but with the zoomed-in $200$~kpc field of view as in the left column of Fig. \ref{fig:xrayzoom}. No filter is added so the actual photon counts can be shown. One can see thermal emissions from the youngest jets and shocks, and their straight or bent morphology. In runs 2030 and 0090, the shape of the latest two or three bow-shocks is visible as well.} \label{fig:xrayhard}
\end{figure*}  

Figure \ref{fig:xrayhard} shows a view in a harder X-ray band, $[10,\,30]$~keV, of the same central $200$~kpc, from the $x$ direction. 
The images are generated with the same aperture/exposure parameters, this time just collecting all generated photons; we \arifII{do not apply} the unsharp-mask filter but show the actual photon counts instead. All objects are rather fainter in this band, with counts of about $100$ in the brightest pixels, as opposed to several thousand in the soft X-ray. Shocks and hot spots are now more clearly visible, except in run 0000. 

We can also observe many features of the jet beams: in run 0000, the beams of the youngest jet and its recollimation shocks are the brightest features, standing out for about $50$~kpc in both directions.  The recollimation shocks appear brighter than the terminal shocks, similar to what was recently observed by \cite{clautice_spectacular_2016} in the (rather favourable) case of 3C 111. Run 0030 is offering perhaps the most interesting insight on the jet kinematics, as the beam is clearly seen to bend at the hot spots in a centrally symmetric S-shape.  The bend resembles the prominent feature seen in the VLA image of M87\footnote{\url{http://images.nrao.edu/57}} by \citet{owen_M87radio_2000}.
The brightest spots are usually the post-shock region immediately after
the terminal parts of the jet beams, offset by a few kiloparsec with respect to the jet itself. Run 2030 shows only one bright HS inside its B6, plus hints of fine ripples and small shocks as observed in Fig. \ref{fig:xrayzoom}, but no clear jet beam.  Finally, run 0090 has one central HS, indicative of a likely FR I morphology, and again a clearly bright B6, all with photon counts in the hundreds, but otherwise relatively featureless.
In this view, the cavities are not outshone by the youngest bow-shock B6, providing a better view than the soft X-ray.  \arif{In other words, the soft and hard X-ray bands have the potential to provide useful complementary perspectives that can help disentangle the complicated shock-cavity structures.}

Comparing the $400$ and the $200$~kpc views, we conclude that a nearby cluster offering a higher resolution view of its core presents some notable advantages, mainly regarding the youngest cavities. In the soft X-rays, a well-resolved centre can reveal a fine-structure of weak shocks and ripples at the boundaries of the youngest lobes indicative of its supersonic phase of expansion.
In the 0000 case, very little structure is present in the inner $200$~kpc. Imaging of the cavities outside $200$~kpc can benefit from higher photon counts, but this tells little about the stability of the cool core. By looking at the core with high spatial resolution in hard X-ray, one can most importantly see details of the jet kinematics inaccessible at lower frequencies, and can pin-point where the shocks occur, not necessarily aligned with the jet beams. Also, young lobes are more visible through the bright cocoons, therefore their size can be measured with smaller uncertainties. Since the cluster is fainter in hard X-ray, actual counts can be quite low (a few or a few tens of counts per spaxel) even with the most recent instruments such as NuSTAR (e.g. \citep{wik_bulletcluster_2014}); a nearby object is therefore mandatory. 

\section{Cavity properties and energetics}\label{sec:pdV}
In this section, we quantitatively analyse the effects of jets and bubbles on the energy balance of our model ICM. In order to compute the energy of the cavities, we isolate them in the 3D simulation grid, and evaluate their volume, pressure, and velocity over time. The cavities are, by definition, hotter than soft X-ray gas; we therefore select them using a threshold in the gas temperature
$T>8$~keV. We refer to this selection using the subscript H, for ``Hot'', opposed to X, for ``X-ray gas''. The selection corresponds to the gas highlighted in the temperature rendering of Figs. \ref{fig:0000} to \ref{fig:0090}, but excluding the bow-shocks and the low-temperature core region.

\subsection{Cavity volume}\label{sub:volume}
\begin{figure}
  \resizebox{\hsize}{!}{\includegraphics{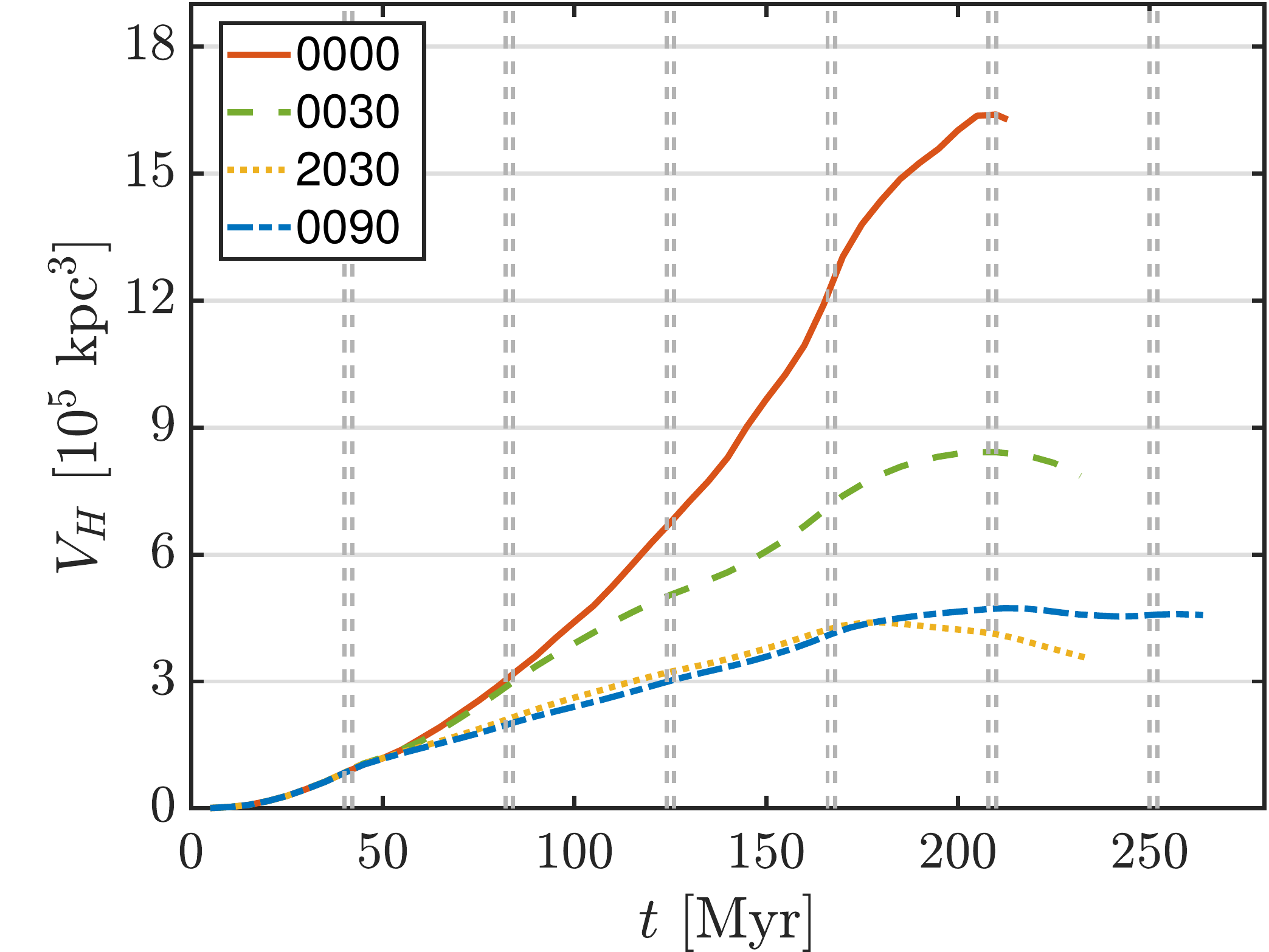}}
  \caption{The actual total volume of the hot ($T>8$~keV) cavities for the jetted runs. The vertical dash-dot grey lines mark the beginning and end of each jet event. Run 0030, and especially 0000, host very large cavity systems, as the jets keep inflating the same bubble complex throughout the whole run. The bubbles in run 0000 are the largest because the jets' momentum has pushed them the furthest away from the cluster core and into a much lower pressure environment. } \label{fig:volume}
\end{figure}

The first important issue that we explore is the volume of the hot cavities (hereafter, referred to as $V_H$), the evolution of which is shown, in units of $10^5$~kpc$^3$, in Fig. \ref{fig:volume}.  \arifII{We emphasise that the plot shows the {\it actual} total volume occupied by {\it all} the cavities within the simulation volume.}  \arifII{The results indicate that jet re-orientation results in a large difference in $V_H$ but in terms of the heating of the ICM, it is not this spread in $V_H$ that is important.   Rather, it is the underlying reason for the spread.   
We noted this in Sect. \ref{sec:zoom}, touch upon it again below, and return to it in Sects. \ref{sub:cavityenergy} and \ref{sec:stability}, where we discuss the implications for the heating of the ICM globally and within the cool core.}

After the first jet, identical in all runs, a clear trend appears: as the orientation of the jets becomes more isotropic, the \arifII{total volume occupied by the cavities declines.}
The two single cavities inflated in run 0000 are by far the largest. \arif{This is partly because all of the jets pump energy into the same cavity and partly because the thrust from all the jets has pushed the cavities further away from the halo centre, where the external pressure of the X-ray gas is lower, enabling the cavities to expand to a larger size than they would have had they remained closer to the cluster centre.}
As we have noted previously, this configuration may offer an explanation for the gigantic bubbles and cocoon shock observed in Hydra A~\citep{nulsen_hydraA_2005,wise_hydraA_2007} and MS0735.6+7421~\citep{mcnamara_heatingnature_2005,mcnamara_ultramassiveAGN_2009}.

At the opposite end, $V_H$ in runs 2030 and 0090, the two with the most isotropic jet/bubble distribution, peaks at a much smaller value of about $4.5\times10^5$~kpc$^3$ around $200$~Myr, then saturates at that point (0090) or even decreases slightly (2030) due to cooling and 
turbulent mixing with the ambient X-ray gas (no hot gas has left the simulation box at this point).  \arifII{The reduced size of the cavity volume is directly the result of all the jets primarily inflating detached cavities within the cluster core. The sizes of these cavities is constrained by the high ambient pressure in the core.  Since the cavity structure in run 0030 is intermediate between that of run 0000 and runs 2030/0090, it is not surprising that the total cavity volume} 
lies in an intermediate position, declining after reaching about $8\times10^5$~kpc$^3$.

\subsection{Cavity energy and mechanical work}\label{sub:cavityenergy}

\arif{
From the pressure, volume, and velocity of all the cells inside the bubbles, we can estimate the bubbles' energy content. By combining this with the total injected jet power, we can estimate the energy transmitted to the intracluster gas, the primary quantity of interest in radio-mode AGN feedback. A direct derivation of this is not possible from the observations and therefore the quantity most often reported and used to estimate the jet power is the enthalpy of the hot gas bubbles \citep{churazov_calorimeter_2002,mcnamara_heating_2007}, which is equivalent to the minimum energy associated with the hot gas bubbles. We therefore compute both the actual energy in the hot gas and \arifII{corresponding proxies based on approximations that observers use.   However, we want to make clear that in the present analysis, the latter are not directly comparable to the quantities deduced from actual observations because we neither account for projection effects nor discount low-contrast bubbles that would be difficult to discern in the X-ray images.  Moreover, we only calculate the total across all the bubbles; that is, we do not distinguish between the individual bubbles even when they are physically disconnected.}
}

The first quantity we consider is the $p\;dV$ work done by the expanding cavities on the surrounding gas to effect its displacement, under the assumptions that the radiative losses suffered are negligible. \arif{This work is defined as follows
\citep{brennen_bubble_1995}.
\begin{equation} \label{eq:wpdv}
{W_X}^{pdV} (t) = \int_{V_H(0)}^{V_H(t)} \left[p_H(\tau)-p_{ICM, shell}(\tau)\right]\ {d V_H(\tau)}
.\end{equation}
Here, $p_H$ is the pressure inside the expanding hot bubbles and $p_{ICM, shell}$ is the pressure of the ambient gas at the boundary of the bubbles, which we neglect under the assumption that $p_H \gg P_{ICM, shell}$.}   
In observed clusters, the time evolution is not available for the integration, therefore  the proxy,
\begin{equation} 
{W'_X}^{pdV} = p_H V_H
,\end{equation}
is often used, where $p_H$ and $V_H$ are estimated directly from the X-ray data under the assumption that the observed bubbles are effectively in pressure equilibrium with their surroundings and therefore, $p_H \approx P_{ICM, shell}$ \citep{churazov_calorimeter_2002,mcnamara_heating_2007}. 


We also compute the total energy of the hot cavities
\begin{equation} \label{eq:eh}
E_H              =  \frac{1}{\gamma-1} p_H V_H  + T_H
,\end{equation}
where $T_H$ is the kinetic energy of the bubbles. We adopt $\gamma = 5/3$, as we have non-relativistic jets. In observations, $T_H$ is often neglected/undetectable, and therefore the cavity energy is approximated using its thermal component only:
\begin{equation}
E_H' = \frac{1}{\gamma-1} p_H V_H.
\end{equation}
However, the contribution of $T_H$ (which includes both ordered and turbulent motions) is often not negligible.  All our cavities show velocity structure on both large (vortex ring) and small scales, including a high degree of \arif{ turbulence both internal to the cavity and at its boundary. 
\citet{cielo_3d_2014} quantified the turbulent energy alone to be about $20\%$ of the thermal energy in similar cavities.}

The quantities ${W'_X}^{pdV}$ and $E'_H$ are useful in producing an estimate of the \arif{(minimum)} total jet power:
\begin{equation}\label{eq:ejet'}
E_{jet}' = E_H' + {W'_X}^{pdV}  = \frac{\gamma}{\gamma-1} p_H V_H. 
\end{equation}
In the simulations, however, the exact value $E_{jet}$ is of course known exactly: 
\begin{equation} \label{eq:ejet}
E_{jet}(t)=\int_{0}^t \left[ P_{jet}(\tau)+U_{jet}(\tau)\right] \ d\tau,
\end{equation}
where $P_{jet}+U_{jet}=1.2\times10^{45}$~erg/s when jets are on and zero otherwise, as seen in Equation \ref{eq:Ujet}.

\begin{figure}
  \resizebox{\hsize}{!}{\includegraphics{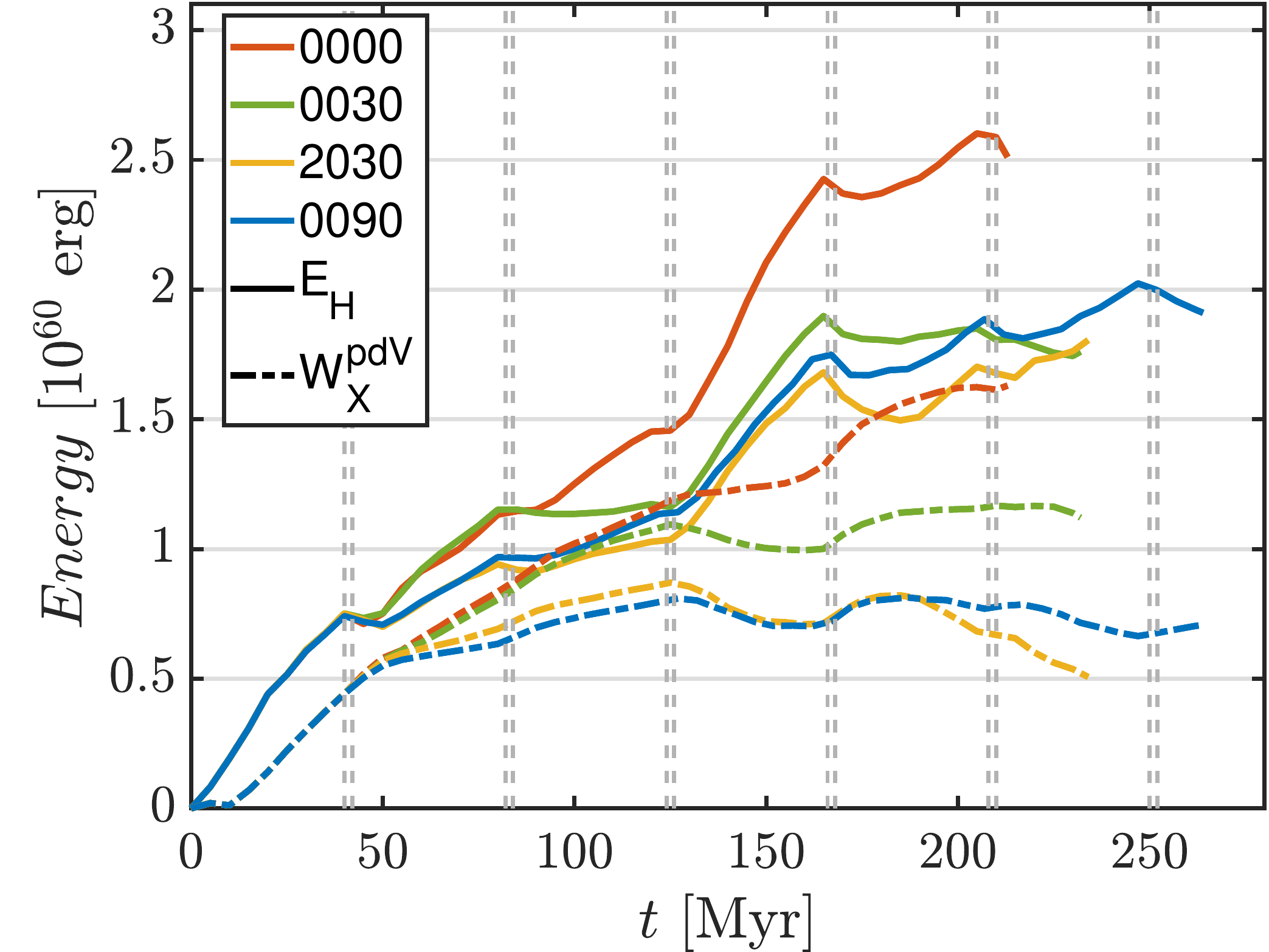}}
  \resizebox{\hsize}{!}{\includegraphics{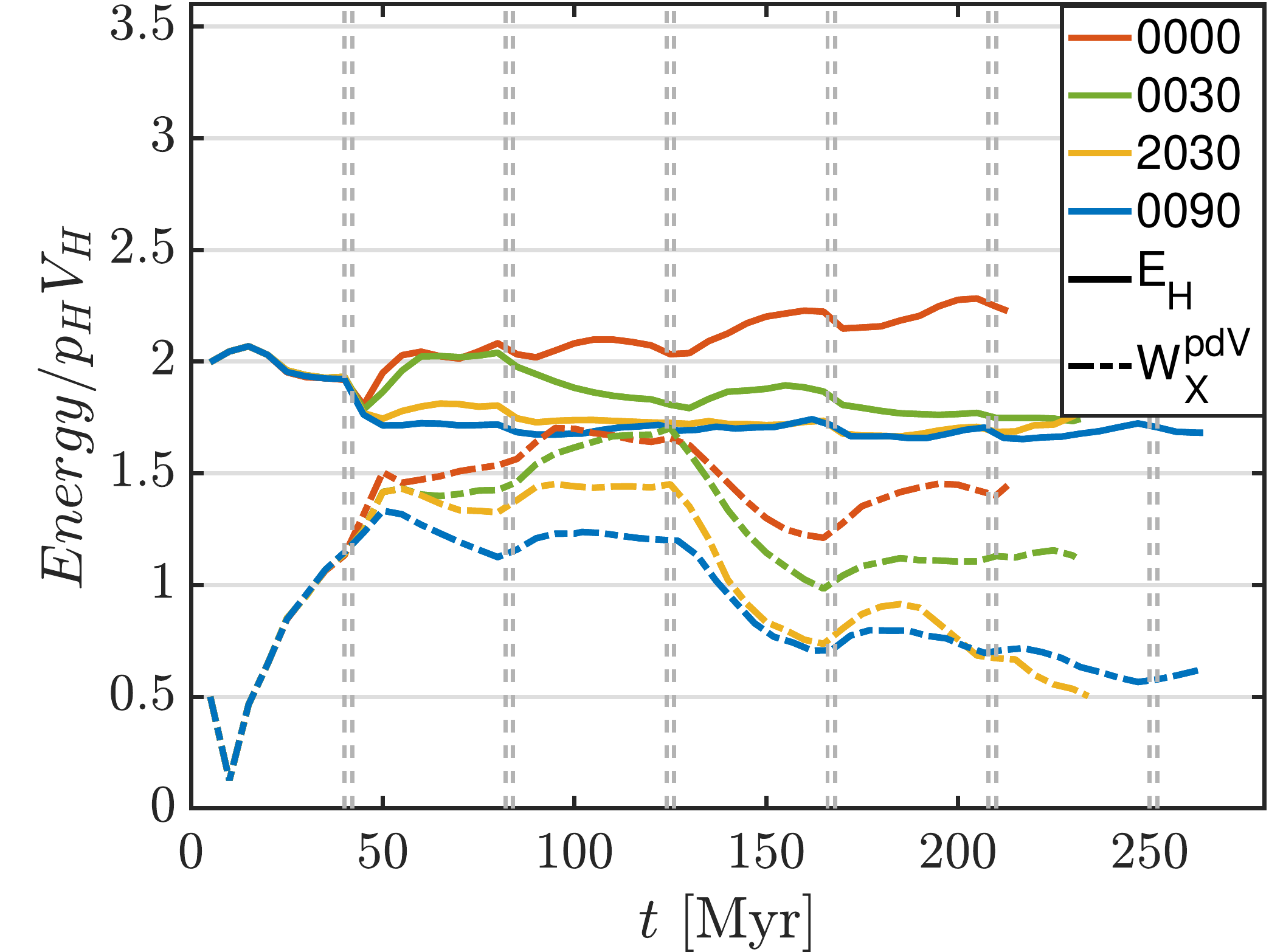}}
  \caption{Total energy $E_H$ of the cavities (solid lines) and mechanical expansion work $W^{pdV}_X$ (dash-dot lines), in units of $10^{60}$~erg/s (top panel) and scaled to the cavity pressure-volume product $p_H V_H$ (bottom panel).  The vertical dash-dot grey lines mark the beginning and end of each jet. Run 0000 shows the highest energies.} 
  \label{fig:energies}
\end{figure}

Figure \ref{fig:energies} shows the evolution of {$E_H$ (solid line) and $W^{pdV}_X$ (dash-dot line)} in units of $10^{60}$~erg (top panel) and scaled to $p_H V_H$ (bottom panel). 

\arif{Focusing first on the top panel of Fig. \ref{fig:energies}, we find that the total energy of the hot cavities, $E_H$, in run 0000 (the one with the fixed-axis jets)
increases almost linearly with time, modulo minor hiccups corresponding to the end of each successive jet event, until the end of the third jet. Then $E_H$ rises very steeply during the fourth jet event and then returns to a more modest rise during the fifth cycle.

In run 0030, $E_H$ tracks the increase in run 0000 over the course of the first two jet cycles, and then flattens to a plateau at about $10^{60}$~erg/s during the third jet, before rising steeply during the fourth cycle.  It again becomes flat during the fifth cycle.

The total cavity energy in runs 2030 and 0090 (the two with maximally re-orientating jets) behaves similarly.  They start like all the others but then grow less steeply over the course of the second and the third jet cycles.  During the fourth cycle, $E_H$ rises very steeply, as in the previous two cases, and then settles back into the slower but steady increase, albeit with fluctuations.

The detailed nature of $E_H(t)$ across the different runs depends sensitively on the extent to which the jets are inflating fully detached cavities, or flowing into pre-existing cavities, or being deflected, and even whether the inflating cavity breaks into a nearby pre-existing cavity.   However, there are two general evolutionary trends that stand out:
\begin{enumerate}
\item{} The overall behaviour of $E_H$ in run 0000 (the one with fixed-axis jets) is different from that of runs 2030 and 0090 (the two with maximally re-orientating jets).  That run 0000
has the largest $E_H$ overall is not entirely surprising since all the jets vent into the same two cavities, losing little energy along the way.   We have already shown that the resulting cavities have the largest total volume; this allows for, as we will show shortly, large-scale circulation flows and turbulence, resulting in higher kinetic energy, $T_H$, within the two main bubbles.  
\item{} The fact that all four runs exhibit a steep rise during the fourth jet cycle indicates that some effect, other than jet history, cuts across all four simulations.   This, it turns out, is the result of the slight contraction of the gas distribution (see also Sect. \ref{sec:stability}) due to cooling.  This particular jet cycle roughly corresponds to one core cooling time for the initial gas distribution.
\end{enumerate}
}
\sI{We note that \citet{mendygral_syntheticAGN_2011}
 performed an analysis of their simulated X-ray data, based on an approach mimicking that used in observations, for the case of non-reorienting jets, and found that the difference between their results for cavity enthalpy and the actual value is a factor of 2.\LEt{Please check that I have retained your intended meaning.}}


\arif{
When looking at $E_H/\left(p_H V_H\right)$ (lower panel of Fig. \ref{fig:energies}), we see very mild time evolution.  On the whole, $E_H\approx \left( 2\pm 0.3 \right) p_H V_H$ summarises the results of all of our runs.  The thermal energy of the bubbles, taken alone, would yield a constant $1/\left(\gamma-1\right) = 1.5$ in this diagram; the difference is due to the kinetic component $T_H$.  Here we can clearly see that the curve for run 0000 is slightly above $2$ and is gently rising with time.   In other words, the kinetic and turbulent energy in the two large cavities continues to build up with time, as we indicated above.  More generally, we see that over the first jet event, the kinetic component comprises $\sim 25\%$ of $E_H$. Thereafter, in all but run 0000, it eventually converges to $\sim 11\%$.  This happens more gradually in run 0030, while in runs 2030 and 0090 (where the new bubbles are completely unconnected), the drop is sudden and happens immediately after the first jet cycle.}

\arif{The mechanical expansion work $W_X^{pdV}$ (upper panel) grows steeply during the first jet event.  Thereafter, in the runs in which the jets often spill into pre-existing cavities (0000, 0030), the bubbles tend to go further out and grow larger, while in the runs with significant re-orientation (2030 and 0090), the mechanical expansion work saturates to a lower value immediately after $\sim 150-200$~Myrs.  This behaviour is a consequence of $V_H$ reaching a constant value in Fig. \ref{fig:volume}.} 
The fact that both $E_H$ and $W_X^{pdV}$ are saturating well before the end of our runs suggests that we have reached a steady state.  Additionally, we want to emphasise that although $W_X^{pdV}$ is the smallest in run 2030 and 0090, the expansion work in those runs is performed in the core (as we noted in Sect. \ref{sub:2030} and \ref{sub:0090}), rather than 
in the outskirts.  
\arifII{This is critical since most of the radiative losses occur in the cluster core.}

\arif{As a final note, we draw attention to the fact that the value of 
$W_X^{pdV}$ is \emph{not} proportional to the product $p_H V_H$; on the other hand, $E_H$ is, even when including the kinetic contribution $T_H$.}

\subsection{Global ICM heating/cooling balance}\label{sub:wmax}

The expansion work $W_X^{pdV}$ represents only part of the energy deposited into the X-ray gas over time; a large part of the ICM heating by the AGN is expected to happen through dissipation of the secondary shocks, velocities, and waves generated by the jets and by the bubbles as they rise and expand in the ICM at the expense of $E_H$.  This is pointed out for example by \cite{nusser_suppressing_2006}, who also noted that the compression of the ICM gas by both the expansion of the bubbles and by shocks can lead to enhanced radiative losses.  We indicate these extra cooling losses as $W_{rad}$ in our energy balance calculation. This term is, however, negligible in most of our simulations (and in any case is much smaller, in absolute value, than the reduction of the global cooling luminosity of the halo due to the AGN feedback) as we show in Sect. \ref{sub:cooling}.

We can now introduce a global upper limit on the ICM heating $W_{max}$, defined as all the jet energy that is neither ``stored'' in the cavities nor radiated away; that is, energy that in some way has been transmitted to the X-ray gas:
\begin{equation}\label{eqa:Wmax}
W_{max} = E_{jet} - E_H - W_{rad}.
\end{equation}
$W_{max}$  thus includes, besides the expansion work, all weak shocks, the ripples, the turbulence and the waves that develop. The dissipation of these features into heat is not instantaneous, so a fraction of energy may still remain in mechanical form rather than internal energy, hence the upper limit character of $W_{max}$. A distinction of the different processes will be the subject of paper II.

\begin{figure}
\resizebox{\hsize}{!}{\includegraphics{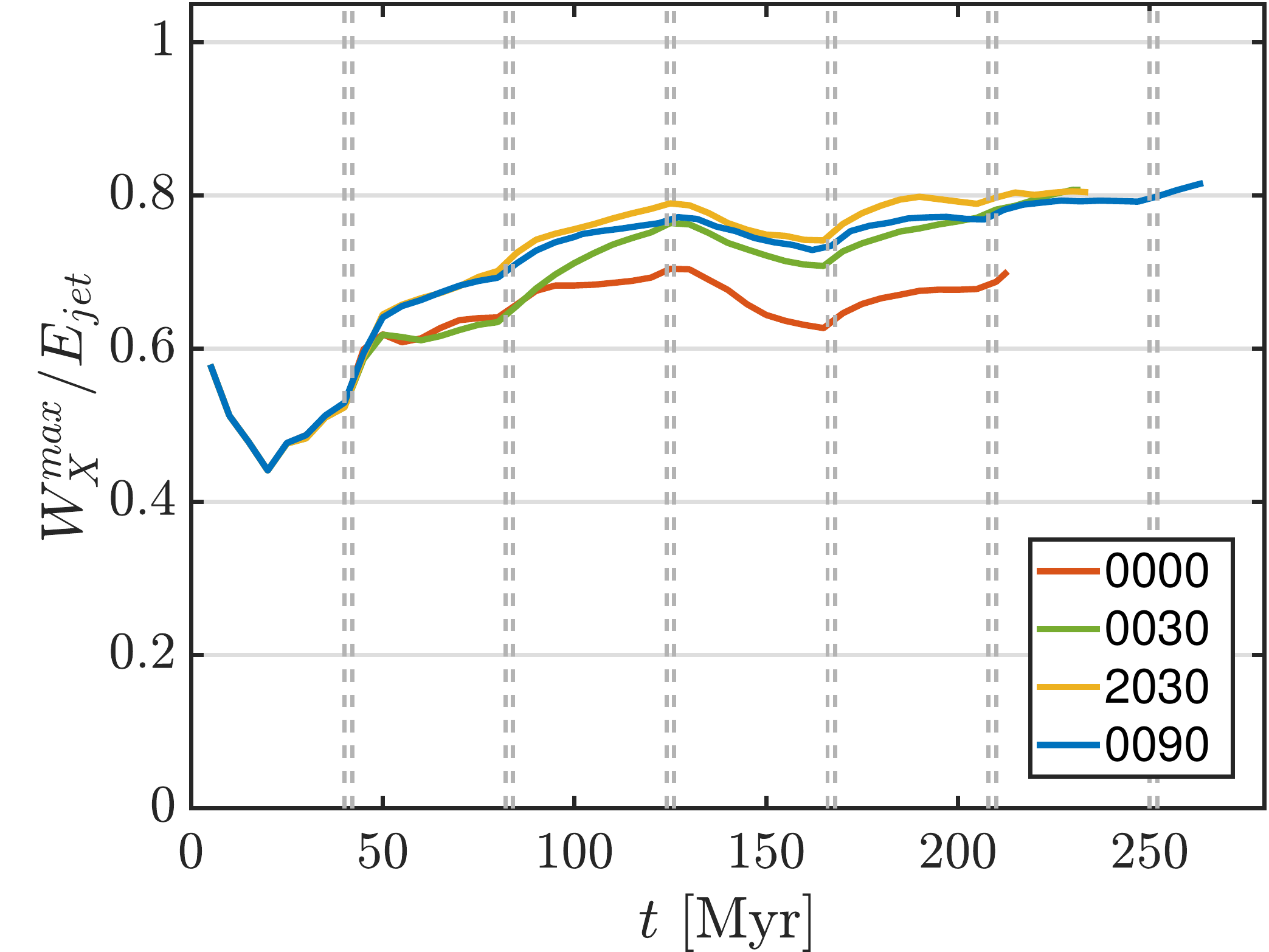}}
\caption{Time evolution of $W_{max}$, upper limit on the {\it global}  amount of energy available to heat the ICM, in units of the jet input energy $E_{jet}$ (Equation \ref{eq:ejet}). The vertical dash-dot grey lines represent the beginning and end of each jet event. Most re-orienting jets converge to a value of about $80\%$ towards the end of the simulation; they all release more energy than the 0000 case, which does not exceed $\sim 65\%$.}
\label{fig:wmax}
\end{figure}

In Fig. \ref{fig:wmax}, we plot $W_{max}$ in units of $E_{jet}$ (Eq. \ref{eq:ejet}), that is, the fraction of jet input energy available for heating the X-ray gas. 
The first shock initially expends energy inflating the first cavity; this is why the value starts low, at about $40\%$. In all runs the energy fraction then grows over time, albeit slightly more efficiently for 2030 and 0090. All runs see a drop at about $130$~Myr, at the start of the fourth jet event. After this point, run 0000 ends at $\sim65\%$, while all other runs set around $80\%$, with a slight increasing outlook.

\arifII{From a global heating/cooling balance perspective, the difference between $65\%$ and $80\%$ may not seem very significant.  However, this overall jet-ICM coupling efficiency factor represents only part of the picture. The other, perhaps important, aspect is where the heat is deposited.   When the jet axis is fixed (run 0000), the jet energy flows out of the core and is deposited a couple of hundred kiloparsecs from the cluster centre.  With moderate-to-signficant jet re-orientation (runs 2030 and 0090), the energy is increasingly deposited in the cooling core.}


\subsection{Re-orienting jets and cooling luminosity}\label{sub:cooling}

\begin{figure}
  \resizebox{\hsize}{!}{\includegraphics{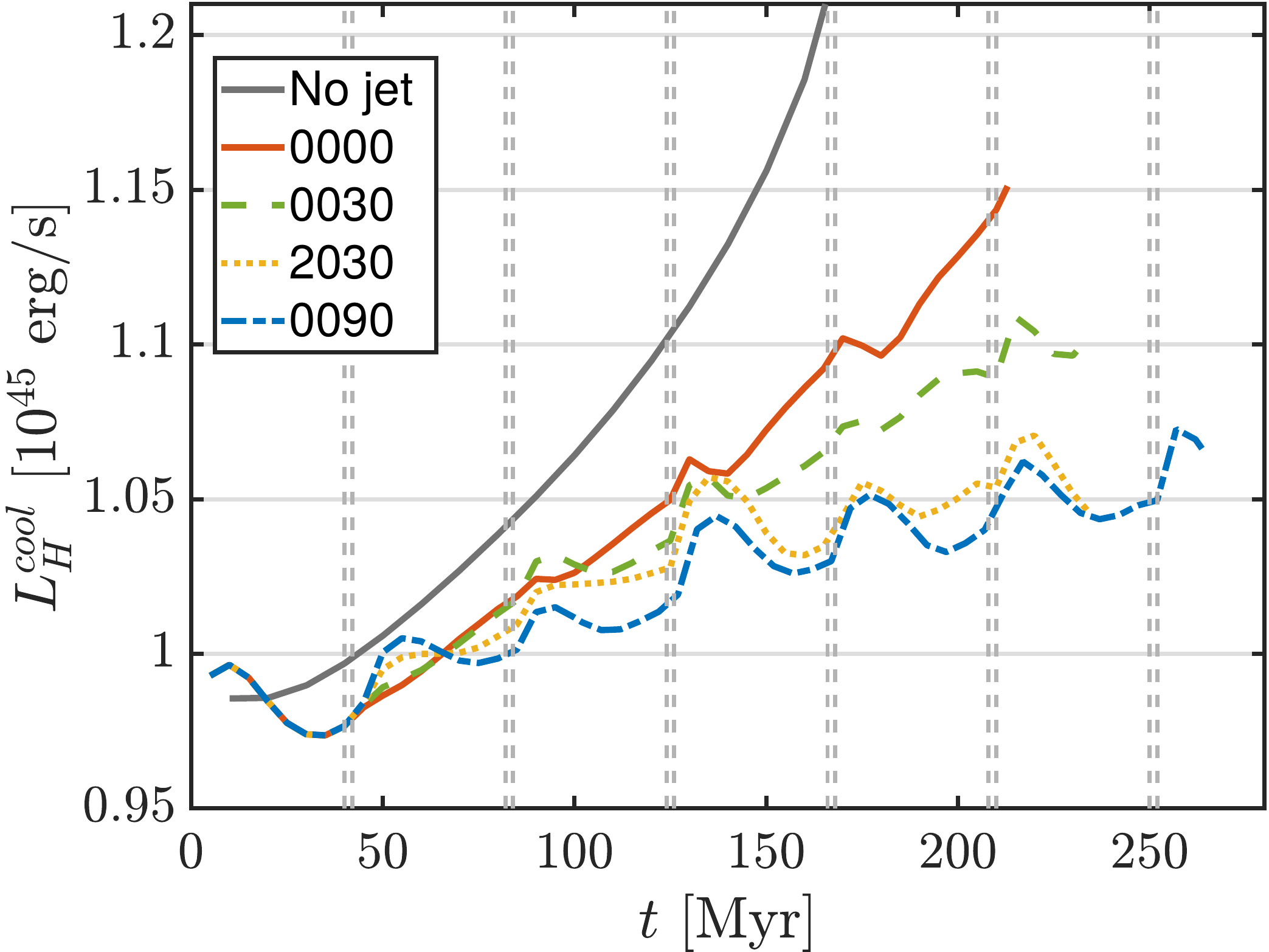}}
  \caption{Halo cooling luminosity within the central $300$~kpc. After the first common jet event, the non-reorienting jets of run 0000 can just delay the cooling catastrophe by $100-200$~Myr, while the more isotropic the jet distribution, the more $L^{cool}_H$ is reduced. All lines show regular ``bumps'' of increased luminosity associated with bow-shocks.  } \label{fig:lcool}
\end{figure}

\arifII{Figure \ref{fig:lcool} shows the evolution of $L^{cool}_H$, the total halo cooling luminosity of the gas within the central $300$~kpc,\footnote{Integrating to a larger radius does not change the cooling significantly.} in units of $10^{45}$~erg/s. 

The grey solid line shows the cooling luminosity of the no-jet \emph{pure cooling} run.  In this case, not only does  $L^{cool}_H$ rise monotonically but the corresponding curve also steepens with time.   
}

\arifII{In comparison to the no-jet run, all jetted runs show 
cooling luminosity that grows more slowly over the course of the simulation.   This is not surprising since heating by the jets stops the ICM gas from growing denser as quickly\LEt{the idea of a change in speed is indicated here already.} as it would in their absence.  The gas still grows denser though, because we have not attempted to tune the jets to compensate fully for the cooling losses.   Still, we highlight that $L^{cool}_H$ grows at different rates in the four jetted runs even though the amount of energy in the jets is the same.

In run 0000, $L^{cool}_H $ grows the fastest.  Specifically, the very first jet episode not only expels gas from the cluster centre (c.f. Section \ref{sub:coregasmass}) but also pushes the distribution out of equilibrium.  As a result, the cooling luminosity declines.  By the end of the event, however, the gas distribution and the cooling luminosity have rebounded (helped along by the backflows and subsequently, cooling flows; see Section \ref{sub:io}) and thereafter, the luminosity increases linearly with time until $\sim 150$~Myrs, after which it increase more rapidly.  This is the case despite the fact that the jets in this run inflate the largest radio cavities, and achieve the highest values for the expansion work $W_X^{pdV}$.  These two outcomes, though seemingly contradictory, can be reconciled by noting that the bubbles in this case are inflated beyond the cluster core.  We can anticipate that this run will experience runaway cooling, similar to the \emph{pure cooling} case and the jets will have only served to delay this by $100-200$~Myrs, in agreement with the findings of several previous numerical studies starting with \citep{vernaleo_problems_2006}.

Looking at run 0000 in more detail, we note small upward ``glitches'' coinciding with the jets.  These glitches reflect enhanced radiative losses ($W_{rad}$ in Equation \ref{eqa:Wmax})
due to compressions induced by the shocks and the cavities.  The fact that the enhancement is not very pronounced is also the result of the jets primarily affecting gas outside the cluster core where the gas is less dense and the cooling less intense.

As the degree of jet re-reorientation increases, the rate at which $L^{cool}_H $ grows is reduced.   In runs 0030 and 2030, $L^{cool}_H $ tracks the cooling luminosity for run 0000 until the end of the second jet cycle, and then the slope of the linear rise becomes slightly shallower, with run 2030 showing a greater flattening than run 0030 and flattening even more after the third jet.   The differences between 0030 and 2030 are due to more efficient heating of the core gas through the inflation of detached cavities in the latter.   All the cavities in run 0090 are detached and form within the cluster core.  Correspondingly, the resulting $L^{cool}_H (t)$ has the shallowest (linear) rise following the first jet cycle.  Interestingly, the late time behaviour of run 2030 appears to approach that of run 0090.

In the run 0090 especially, the \emph{bumps} associated with the jet events are easy to see.   Their larger amplitude is the result of the jets compressing denser, already strongly cooling gas in the cluster core, which also gives rises to the brighter bow-shocks that we observed in Sect. \ref{sec:description} in our synthetic X-ray pictures of the more isotropic jets (Figs. \ref{fig:0000} to \ref{fig:xrayzoom}).


We can use these fluctuations to estimate the $W_{rad}$ term of Eq. \ref{eqa:Wmax}.   Measuring the height of the peaks above the ``valleys'' in the plotted line, we obtain approximately  $\Delta L^{cool}_H \approx 2.5\times10^{43}$~erg/s in the highest peaks. If we then, conservatively, round it to an average of $2\times 10^{43}$~erg/s, constant over time, we get
\begin{equation}
W_{rad} 
\simeq 6 \times 10^{58}\; \mathrm{erg}\; \left(\frac{t}{100\,\mathrm{Myr}}\right)   \simeq 0.018 E_{jet}.
\end{equation}
\arifII{This confirms that $W_{rad}$ is a negligibly small correction to 
$W_{max}$, which is of order $0.8 E_{jet}$ (see Fig. \ref{fig:wmax}) and 
the corresponding $\Delta L^{cool}_H$ is similarly a minor ($\sim 2\%$) correction to the overall cooling luminosity.}

By comparing the cooling rate ($L_H^{cool}$) with the heating ($W_{max}$), we gain insight into the net energy change of the X-ray gas and specifically the heating-cooling balance.  Focusing on our two most isotropic runs, 2030 and 0090, we find that at $250$~Myr, 
\begin{align}
L_H^{cool} & \simeq1.05\times10^{45}\;\rm{erg/s} \nonumber \\
\begin{split}
W_{max} & \simeq 0.8\;E_{jet} =0.8\;\left(\frac{40\;\rm{Myr}}{42\;\rm{Myr}} \right)\; 1.2\times10^{45}\;\rm{erg/s}  \\
                &\simeq 9.14\times10^{44}\;\rm{ erg/s}.
\end{split}\nonumber
\end{align}


We had hinted at this overall heating/cooling imbalance.  Even in the best-case scenarios, the ones with maximally re-orientating jets, a slightly higher value of $P_{jet}+U_{jet} \simeq 1.3\times10^{45}$~erg/s would provide a better \emph{global} heating/cooling balance.  However, it is important to note that the heating-cooling balance is not merely a function of jet power.  Based on what we have observed in the simulations, we expect that, while a higher jet power would stave off runaway cooling for a while longer, in runs 0000 and 0030, for example, catastrophic cooling would eventually set in because none of the jets after the first (and perhaps the second) would actually heat the core.   



\begin{figure*}
  \centering
  \includegraphics[width=17cm]{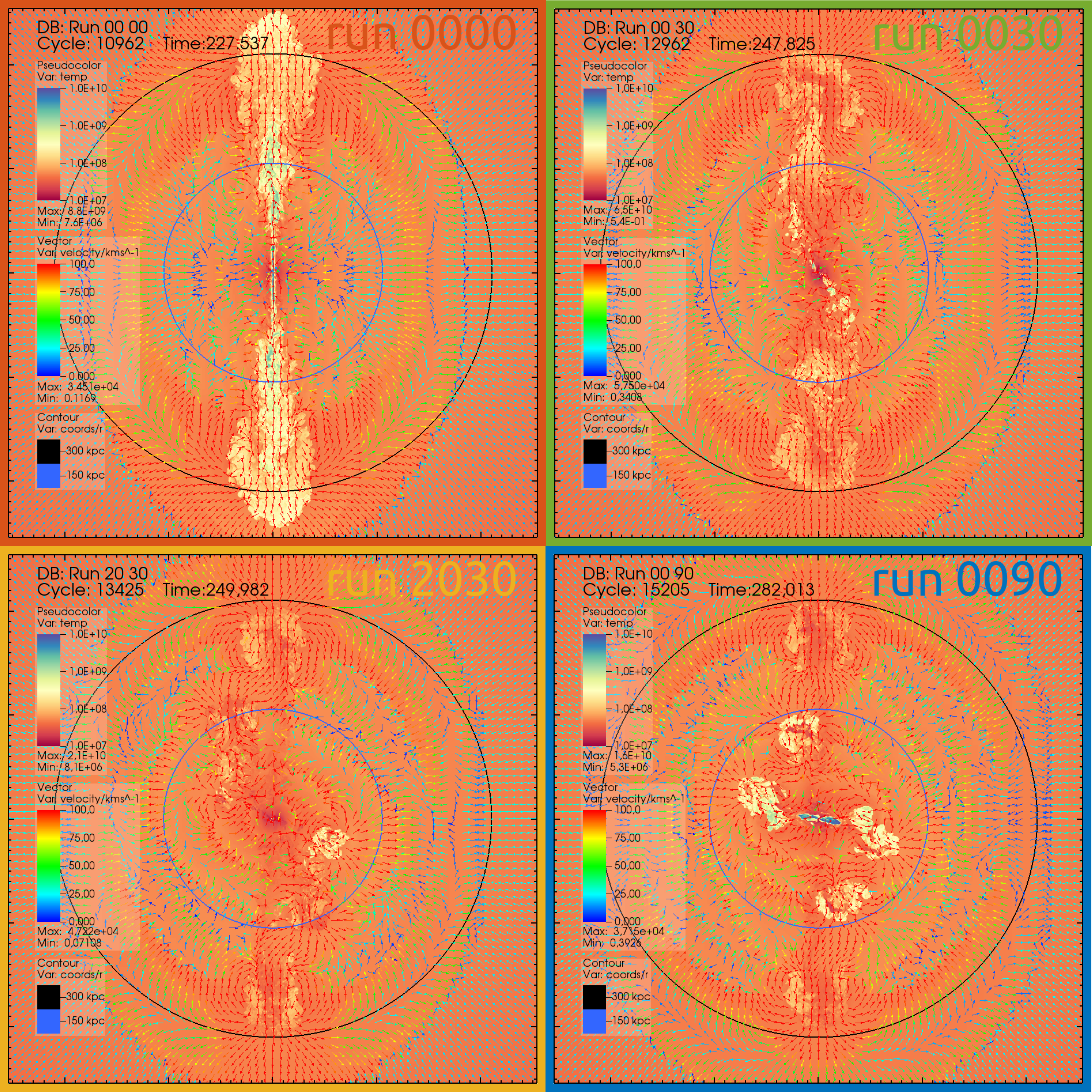}
  \caption{Temperature slices along the $y=0$ plane, showing the final snapshots of all simulations. The superimposed arrows trace the velocity field (capped at $100$~km/s) and spherical contours mark $150$ and $300$~kpc. Fast outflows counteract a slow but steady cooling flow. A complex inflow/outflow pattern develops as bow-shocks propagate in the ICM. However, significant outflows within $150$~kpc are seen almost exclusively in run 2030 and 0090. Hot, sparse and very extended \emph{backflows} wrap around all cavities\LEt{i.e. they encircle them; please verify that this is the intended meaning.}. We note that a knot of cooler gas is present at the cluster centre in all runs.}
  \label{fig:vel}
\end{figure*}


The most important difference between re-orienting and steady jets lies in where the heating is deposited. As noted earlier, the cavities in runs 2030 and 0090 are inflated at smaller radii and therefore the jets deposit the bulk of their energy within the core.  Given that radiative losses occur primarily in the core, stabilising the core is key if AGN feedback is to prevent $L_H^{cool}$ from growing in a runaway fashion.}


\section{Stability of ICM gas}\label{sec:stability}
\subsection{Inflows and outflows}\label{sub:io}

\arifII{
Simulations of AGN feedback in galaxy clusters, in which jets are able to counteract cooling in the cores of CCCs, predict a state of dynamical equilibrium involving a rich pattern of inflows and outflows across a range of spatial scales within the central few hundred kiloparsecs (\citealp{prasad_cool_2015,prasad_cool_2017,yang_how_2016,cielo_backflow_2017,lauetal_perseusar_2017,Prasad_turbulence_clouds_2018,Gaspari_gaskinematics_2018}).  Cooling flows, expanding heated gas, gas displaced by inflating cavities, uplifted gas in the wakes of rising cavities, and backflows all contribute to the velocity structure (c.f. \citealp{neumayer_central_2007}). Gas flows, therefore, provide insight into the interplay between cooling and feedback, and the stability of the cluster core against cooling. In this section, we examine the velocity structure in our simulations. }

\begin{figure}[h!]
  \resizebox{\hsize}{!}{\includegraphics{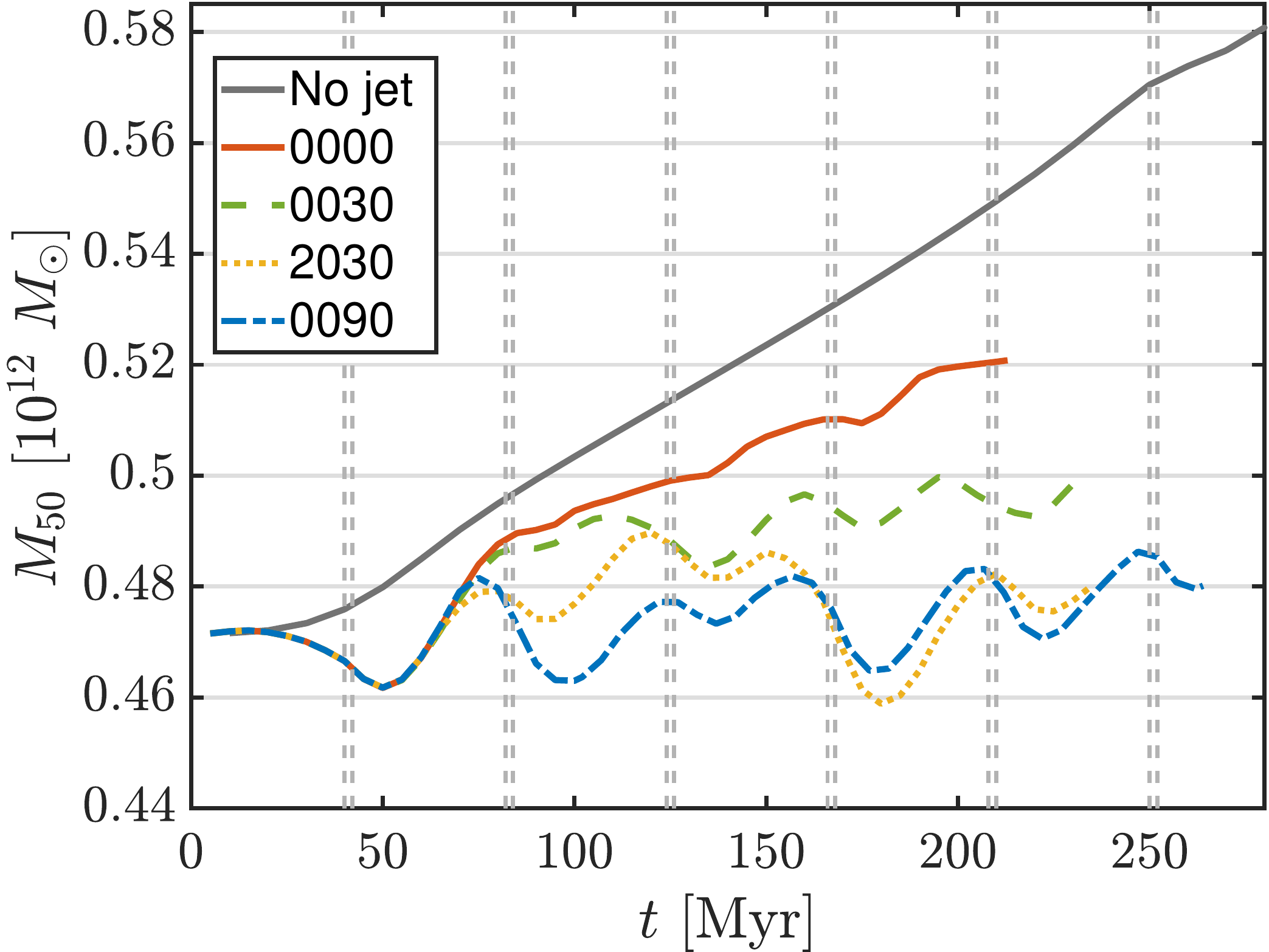}}
  \resizebox{\hsize}{!}{\includegraphics{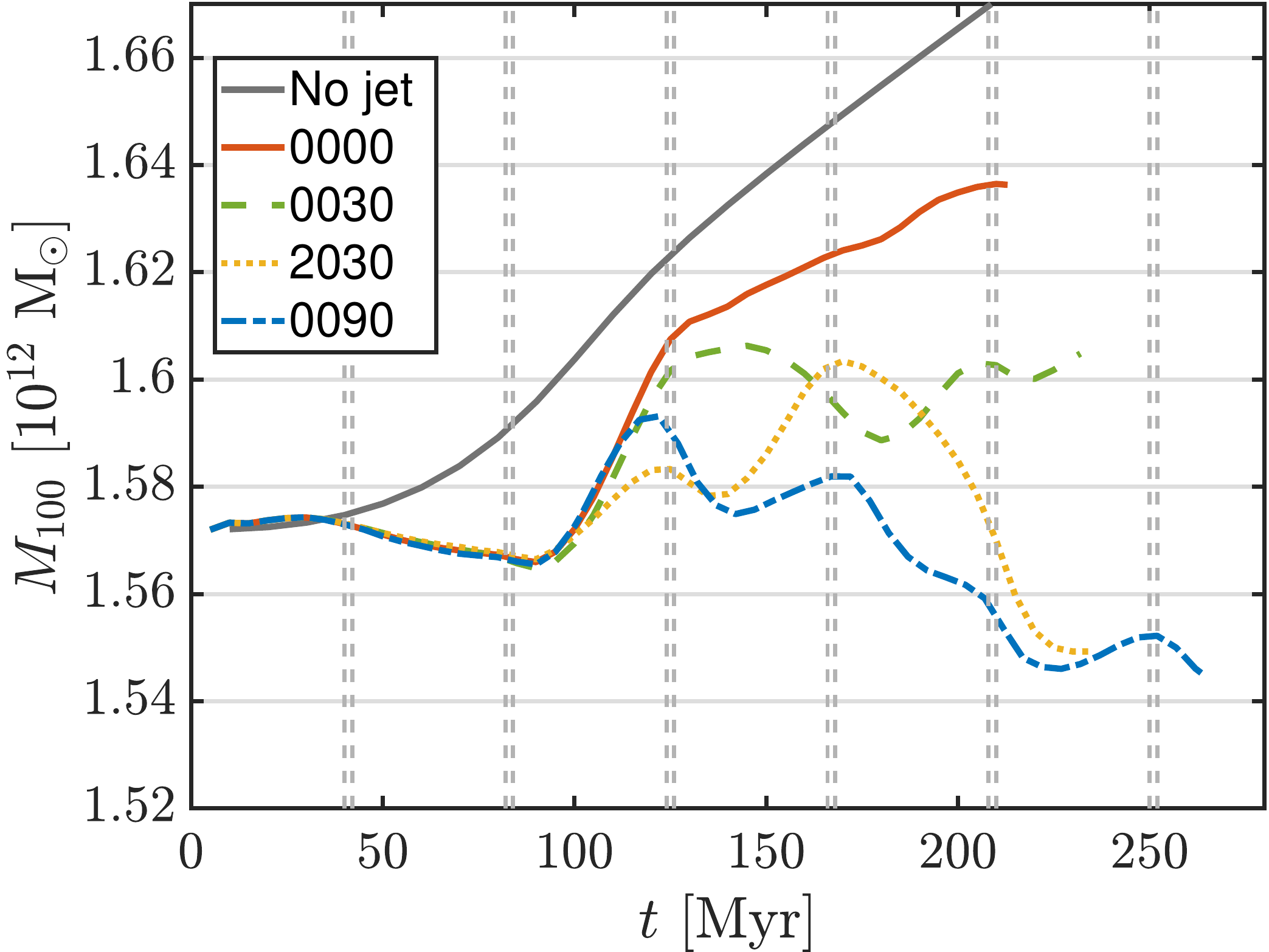}}
  \resizebox{\hsize}{!}{\includegraphics{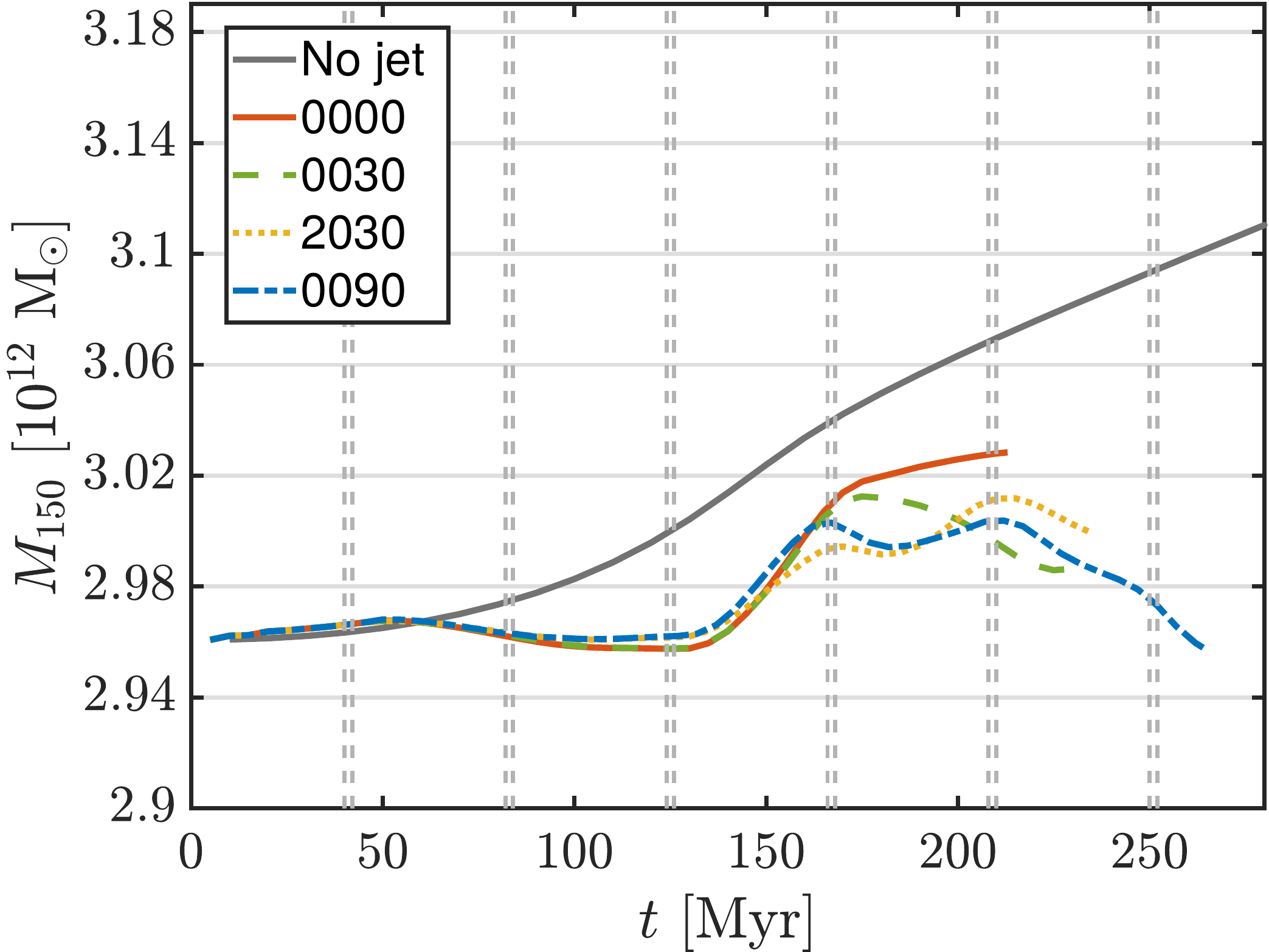}}
  \caption{Evolution of the gas mass $\mathrm{M}$ within the central $50$, $100,$ and $150$~kpc spherical regions, from top to bottom).
  Grey line: pure cooling flow case. Coloured lines: jetted runs, as indicated by the key. Without feedback, $\mathrm{M}_{50}$ increases by more than $20\%$.  For a discussion of the trends exhibited by the jetted runs, 
see the text.}
  \label{fig:massSpheres}
\end{figure}

Figure \ref{fig:vel} shows temperature slices ($400$~kpc a side, along the $y=0$ plane) of our jetted runs, with superimposed arrow plots of the gas velocity (with magnitude capped at $100$~km/s). Two circular contours mark the radii of $150$ and $300$~kpc. These slices are extracted from the same snapshots as Figs. \ref{fig:0000} to \ref{fig:0090}. There are several features in common across the panels:

\arifII{
\begin{enumerate}
\item{} The cavities associated with the first jet have generally reached  (runs 0030, 2030 and 0090), or gone slightly past (run 0000), $300$~kpc and the gas affected by the bow shock ahead of the cavities is flowing outward at $100$~km/s or faster.
\item{} Beyond roughly $300$~kpc from the cluster centre,  the gas is flowing inward at about $25$~km/s in all directions.  This flow is due to cooling of the gas.
\item{} The jets engender a hierarchy of velocity inflows/outflows within a visually easy-to-discern  central ellipsoidal region.  This region roughly corresponds to the expanding cocoon associated with the first jet event, with gas velocities varying over the boundary of the region from $\sim 25-75$~km/s.  The detailed shape and velocity structure of the boundary region depends on the extent of jet re-orientation but all models show this feature. 
\item{} The velocity structure inside this ellipsoidal region comprises (a) large-scale backflow associated with the first jet event, which itself is churned up by outflows and backflows associated with subsequent jets/cavities, the \emph{vortex ring}-like velocity structure of the individual cavities, as well as cooling flows.
\end{enumerate}



Apart from these general features, we draw attention first to run 0000.   In this case, there is a inward velocity flow transverse to the fixed jet axis starting at a distance of approximately $200$~kpc from the cluster centre.   This flow is the product of the backflow amplified by cooling.  The inward velocity varies from $\sim 10$~km/s (blue arrows) to $75$~km/s (yellow arrows) as the flow encounters different cocoons but is always inward flowing.  This flow contributes to the eventual runaway cooling of the cluster core.   In the remaining runs, the large-scale inward flow is increasingly disrupted as larger jet re-orientation angles  result in increasingly isotropic distribution of cavities inflated in the inner $100$~kpc, which in turn lead to the outflows subtending an increasingly larger fraction of solid angle.  However, the central knot of cool ($\sim 10^7$~K) gas is never quite destroyed in any of the cases.



\subsection{The core structure: gas mass}\label{sub:coregasmass}

In light of the above discussion about inflows and outflows,}
in Fig. \ref{fig:massSpheres} we plot the evolution of the gas mass $\mathrm{M}$ within $50$, $100,$ and $150$ kpc (top, middle, bottom panel, respectively) in not only the jetted runs but also, for comparison, in the no-jet, cooling only run.
For the gas distribution to be in dynamical equilibrium, $\mathrm{M}$ ought to converge to a constant value, or perhaps oscillate around one with the same variability as the jets.

\arifII{Not surprisingly, the radiative losses quantified in Fig. \ref{fig:lcool} result in a cooling flow in the no-jet simulation, which in turn leads to monotonically increasing gas mass within the central $50$, $100,$ and $150$~kpc.


In all of the jetted runs, the masses initially follow the pure cooling result until the bow-shock associated with the first jet crosses the volume in question, and the gas is expelled from the volume and out of hydrostatic equilibrium.  Thereafter, the evolution of gas is driven by resettling and cooling, as well as heating and expansion by subsequent jet activity, and of course, the circulation flows that the latter engender.  
}

\begin{figure}
 \resizebox{\hsize}{!}{\includegraphics{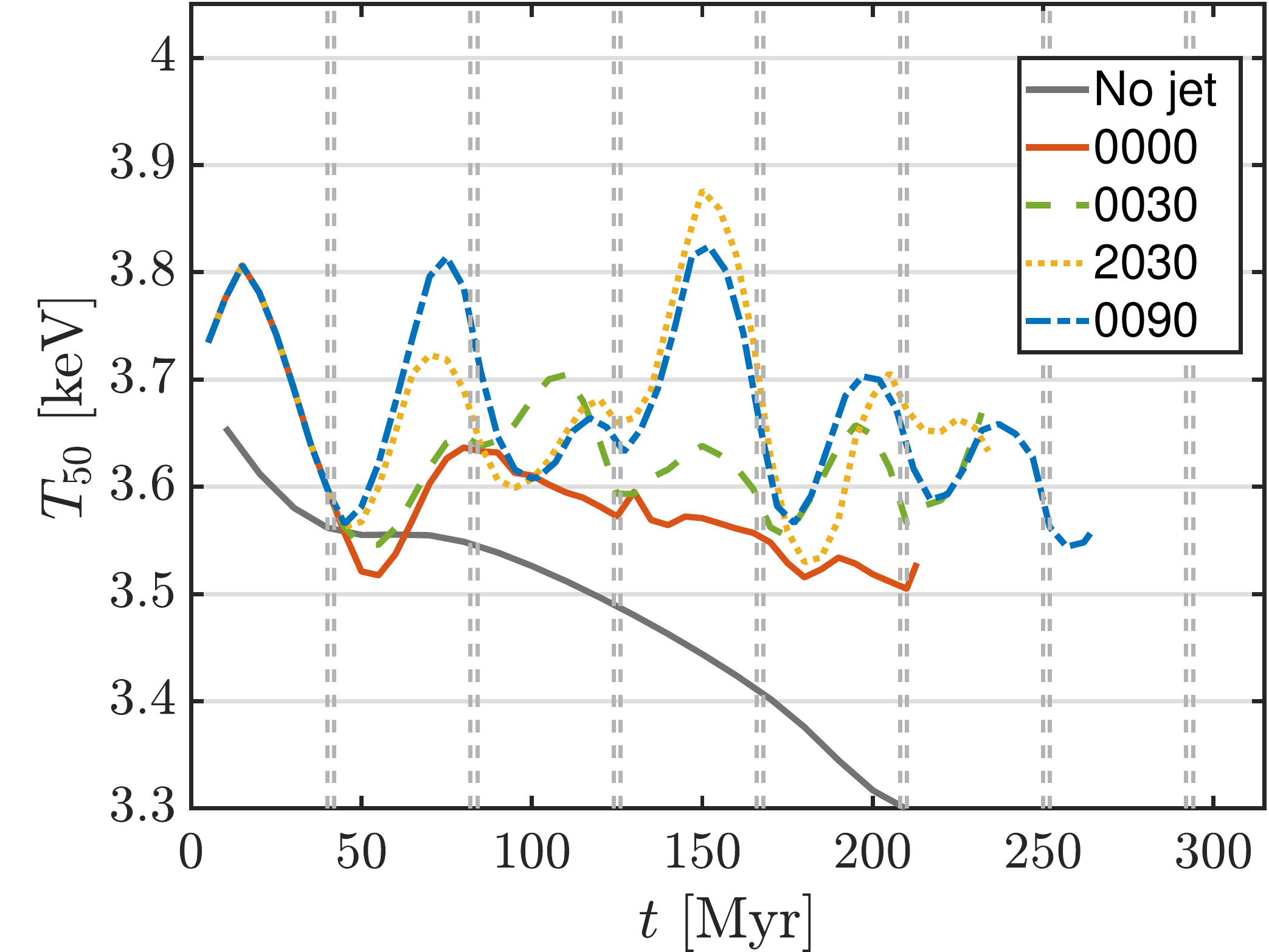}}
 \caption{Evolution of the \emph{mass-averaged} temperature of the ICM in the central $50$~kpc region. The jets temper the temperature decline from the initial value of $3.65$~keV. Oscillations correspond to the bow-shocks. The more isotropic the jets, the gentler the downward trend and 
 more prominent the oscillations.} \label{fig:tcore}
\end{figure}

On the scale of $50$~kpc, the gas masses in all runs initially track each other for $\sim 60$~Myrs, which is roughly the sound crossing time to the edge of the region:   During this time, $M_{50}$ initially drops and then not only rebounds but helped along by the backflows and the transverse cooling flows, overshoots the initial value.  Thereafter, $M_{50}$ in run 0000 rises steadily, albeit less steeply than in the no-jet run.  In runs 2030 and 0090, $M_{50}$ oscillates in response to the jet events, with the detailed character of the oscillations depending on, among other things, whether the specific jet event inflates a new cavity or intersects and channels through an existing one. \arifII{In the mean, the two behave similarly, exhibiting a gentle rise.  In other words, with time, the $M_{50}$ increases, albeit slowly, as anticipated from Fig. \ref{fig:lcool}.  Run 0030, as always, falls in between 0000 and 2030/0090.

Turning to $M_{100}$, all the runs track each other for a time comparable to the sound crossing time. Thereafter, the masses in run 0000 rise more steeply than the pure cooling case at first, and then flattens slightly to continue rising, less rapidly than the pure cooling case but rising steadily nonetheless.  In runs 0030 and 0090, the ones with the least and the most re-orientating jets, respectively, $M_{100}$ rebounds steeply, following the rise in run 0000 until approximately the end of the third jet cycle. Thereafter, $M_{100}$ in run 0030 appears to stabilise, oscillating about a constant value.   Since this happens close to the end of the simulation, we cannot say whether $M_{100}$ will continue to oscillate with each successive jet episode or will eventually start rising again.   Based on the trajectory of the cooling luminosity, we expect the latter.  In run 0090, $M_{100}$ executes mild oscillations but the overall trend is downward.  By the end of the simulation, it is below the starting value.  Run 2030 is in between runs 0030 and 0090: $M_{100}$ decouples from the rest slightly earlier and swings up and down with an amplitude that grows with time,  ending up below the initial value when the simulation stops. 

As for $M_{150}$, all the runs follow each other until roughly the end of the fourth jet episode.  Thereafter, the curve for run 0000 continues to rise.   Once it decouples, $M_{150}$ in run 0030 appears to decline but we cannot tell whether this is a long-term trend or simply a downward part of an oscillation.  In run 0090, the mass within $150$~kpc hovers at a constant value before trending downward to the end of the simulation.  Once again, run 2030 is intermediate between runs 0030 and 0090.

Having considered the trajectories of $M_{50}$, $M_{100}$ and $M_{150}$ individually, we conclude with some global observations:
\begin{enumerate}
\item Even though the jets are injecting the same amount of energy in all four jetted simulations, they clearly cannot thwart cooling  flows in run 0000, the one with the fixed-axis jets.   After the gas in the three regions has reacted to the initial jet events, the corresponding mass begins to rise steadily.  This agrees with our previous observations, and those of 
\citet{vernaleo_problems_2006}, that in due course the jet energy is not only channelled away from the cluster core region but is also transferred to the gas highly anisotropically.  The gas in the core cannot avoid cooling, and the growth of $M_{50}$, $M_{100}$ and $M_{150}$ is the consequence of the transverse circulation-augmented cooling flow apparent in Fig. \ref{fig:vel}.
\item The curves in Fig. \ref{fig:massSpheres} provide strong evidence that the jets in run 0090 have the strongest impact on the gas between $50$ and $150$~kpc.  Heating by the jets is clearly causing the gas in these regions to expand and hence, $M_{100}$ and $M_{150}$ decrease steadily over the long term.  The jets, however, do not appear to be as efficient at preventing the mass within $50$~kpc from growing.  The flow of gas into this region is largely due to backflows but as the gas mass increases, so does the corresponding cooling luminosity. 

It is possible that over time, the combined effect of the outflow that is driving gas out of the central $100$~kpc and the cooling flow in the inner regions will reduce the density of the X-ray gas to a point where it cannot cool efficiently, and thereafter the central region stabilises.   If this is the unfolding trajectory, the likely timescale is $\sim 1$~Gyr, much longer than our simulation run time.   

An alternative approach to offsetting the backflows and quenching the central cooling flow is to ensure that the jets are able to expel an equal amount or more gas from within $50$~kpc.  As it stands, it appears that in the absence of any other effect,  runaway cooling will eventually set in.  Previously, we had suggested that based on the global analysis of heating and cooling, a slight increase in the jet power might do the trick.  However, the results discussed above suggest that apart from boosting the jet power, the jet profile may need to be adjusted to ensure improved coupling between the jets and the gas in the very central region  ($<50$~kpc).  
\sI{Lowering the jets' momentum flux while increasing (or holding constant) their energy flux could do this.}
\end{enumerate}

\subsection{The core structure: gas temperature and thermal energy}\label{sub:coregastemp}

In light of the evolution of the gas mass in the central $50$~kpc, we examine the temperature and the thermal energy of this central region.}
In Fig. \ref{fig:tcore}, we plot $T_{50}$, the (mass-averaged) temperature within $50$~kpc (refer to Fig. \ref{fig:vel} for a visual comparison). Including or leaving out the jet beams/bubbles has little impact on the temperature determination since these are low-density structures. 

From the initial temperature of approximately $3.65$~keV, the no-jet run shows a steady decline, interrupted only by a small plateau around $50$~Myr.  This plateau arises because of the initial rapid cooling rate in the core; the resulting loss of pressure support causes the bulk of the gas within $50$~kpc to flow inwards sufficiently quickly that adiabatic heating due to contraction temporarily balances radiative cooling.  Eventually, however, cooling resumes and dominates.

In the jetted runs, \arifII{the first jet heats and expels the gas from the region.  The gas mass goes down (see Fig. \ref{fig:massSpheres}) and the expansion also drives the gas temperature down.  At this point, the temperature drop is due to more or less adiabatic expansion.
In run 0000, the expansion stalls at $\sim 50$~Myrs. The gas starts to fall back, and as it does so, it experiences compressive heating.  The increase in temperature is tracked by the increase in $M_{50}$. At about $\sim 75$~Myrs, radiative cooling sets in: The temperature begins a steady decline but this time, the mass does not follow.  Instead, $M_{50}$ continues to rise.   We note that during this phase, the rate of decline of $T_{50}$ is slightly shallower than in the pure cooling case because a small fraction of energy from subsequent jet energy dissipates in the region.   

In the re-orientating jet runs, the rises and falls in the mass curves due to the second and subsequent jets are largely replicated in temperature.} The variations are typically $\pm 3-5\%$ and vary from one jet event to another depending on the whole reorientation history.  The oscillations are generally larger in runs 2030 and 0090 since their shocks form within the region. 
\arifII{
On the whole, however, the temperature profiles show a declining trend, with runs 2030 and 0090 exhibiting the gentlest decline while the decline in run 0030 is, as expected, intermediate between these two and run 0000. 
}



\begin{figure}
  \resizebox{\hsize}{!}{\includegraphics{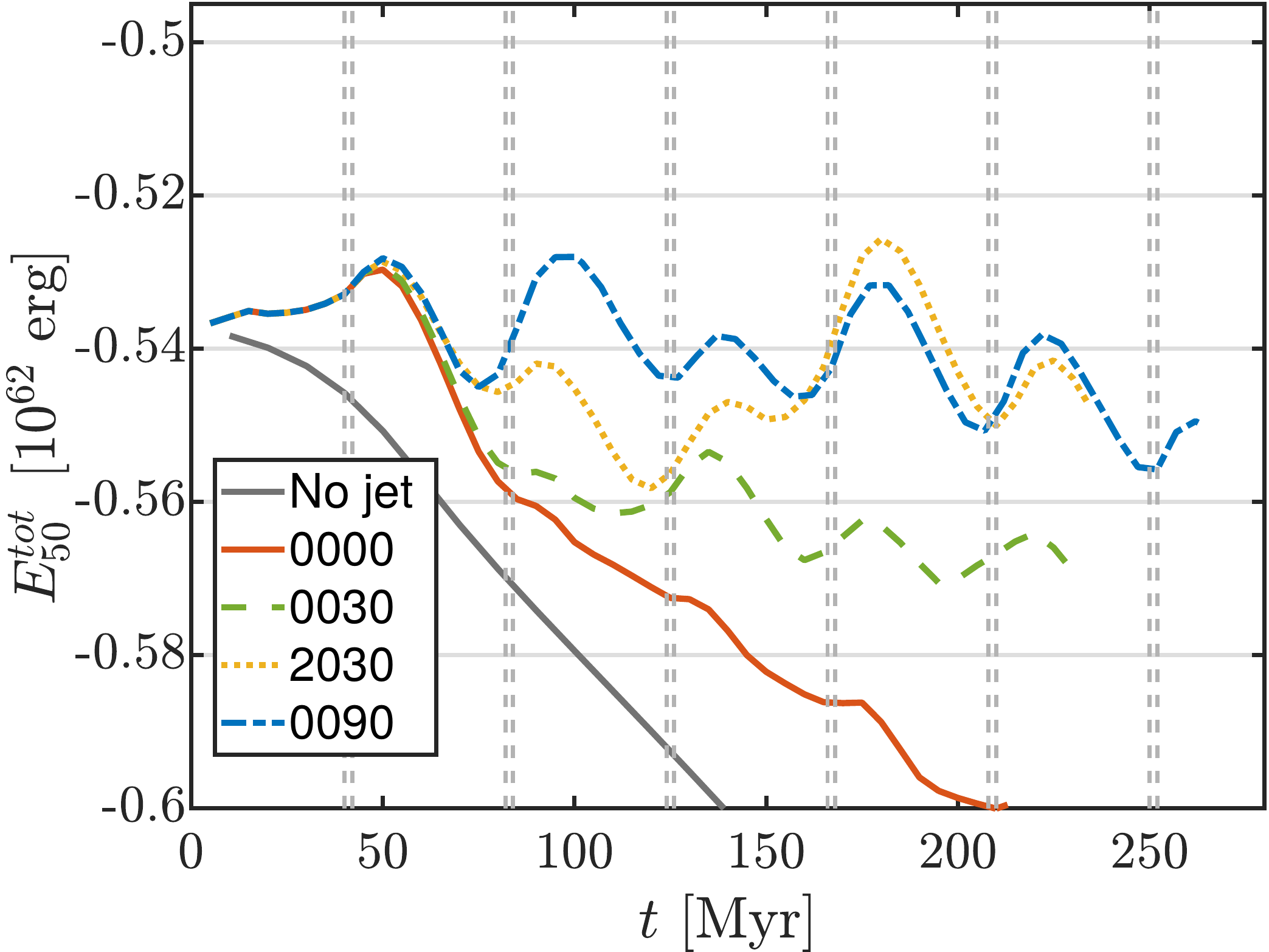}}
  \resizebox{\hsize}{!}{\includegraphics{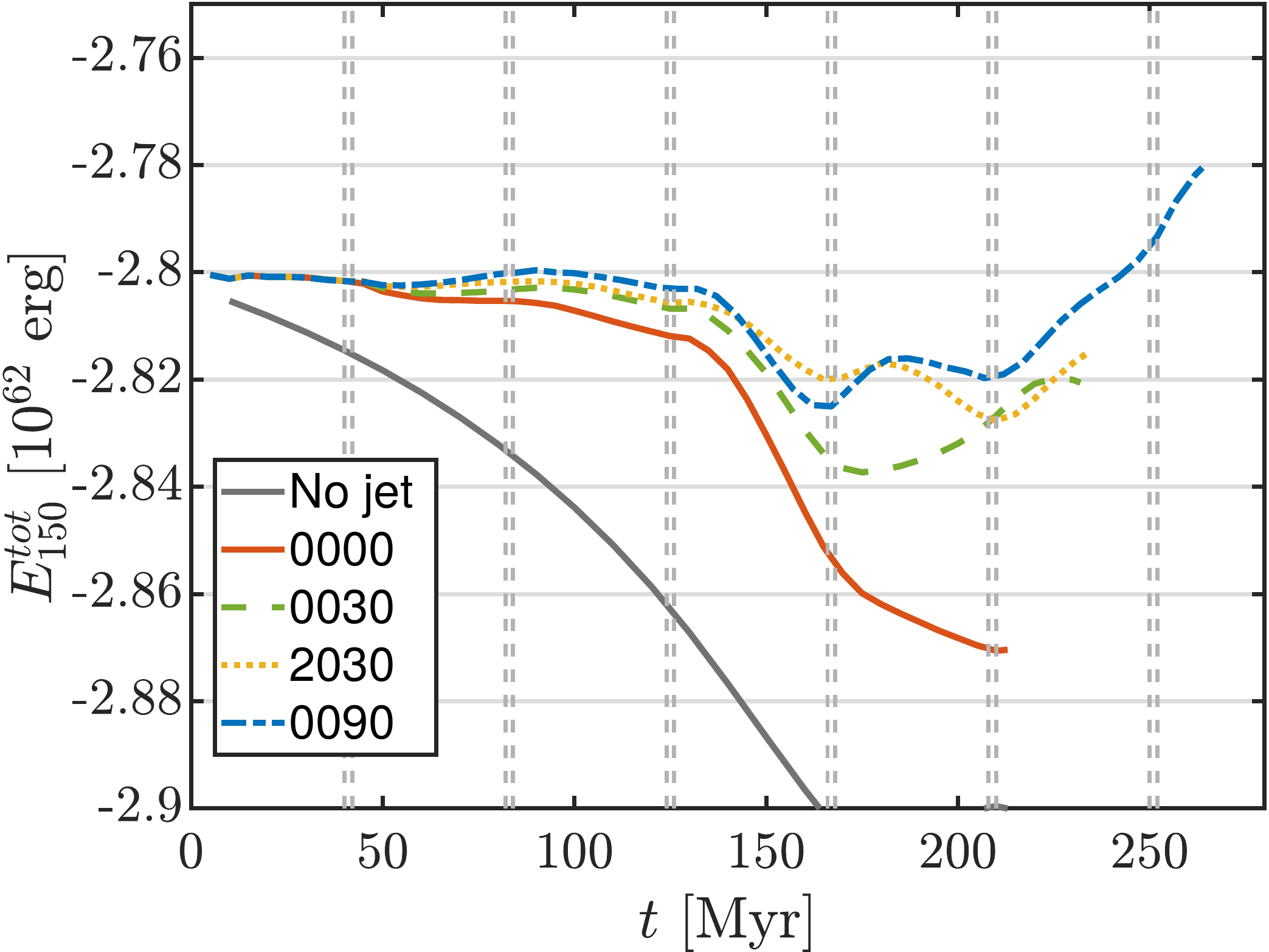}}
  \caption{The time evolution of the total (kinetic + potential + thermal) energy, $E^{tot}$, of the X-ray gas within the central $50$ (top) and $150$~(bottom)~kpc regions.  See the text for a discussion of the trends.
  } \label{fig:etot}
\end{figure}

Similar but smaller amplitude oscillations appear also in the top panel of Fig. \ref{fig:etot}, where we show the evolution of  $E^{tot}_{X50}$ and $E^{tot}_{X150}$, that is, the total (potential + kinetic + thermal) of the X-ray gas only within $50$ and $150$~kpc, respectively. Following \cite{mccarthy_towards_2008}, we choose the gas' total energy as the best indicator for the state of the gaseous halo, as the exchanges between potential and thermal energy are frequent when radio mode AGN feedback is involved. We note that $E^{tot}$ is always a negative quantity, as the gas volume under consideration is gravitationally bound.

As gas cools and flows inwards, both the loss of energy to radiative cooling and the increase in mass drive $E_X^{tot}$ downwards. Injection of thermal and kinetic/turbulent energy, and gas expansion, do the opposite.
Within the $50$~kpc volume, $E_X^{tot}$, like $T_{50}$, oscillates while exhibiting an overall decline. The decline is the steepest in run 0000 and gentle in 2030 and 0090 for reasons already discussed.  The panel, not surprisingly, looks like the mirror of the $M_{50}$ panel.

\arifII{Similarly, the $E_{X150}^{tot}$ panel resembles the mirror of $M_{150}$, mostly, with declining mass corresponding to rising energy and vice versa. There is, however, one small difference: between approximately $50$~Myrs and $130$~Myrs, the mass shows a gentle decline while   $E_{X150}^{tot}$ remains mostly flat in run 0090 but otherwise declines.  The reason, we suspect, is due to radiative cooling. We also note that $E_{X150}^{tot}$ for run 0090 is clearly rising towards the end of the simulation, for the reasons already mentioned when we discussed the mass curves.
}


\section{Discussion}\label{sec:discussion}

\subsection{Comparison with previous numerical work}\label{sub:prevSim}

The interplay between multiple fixed-axis jets and a cooling ICM in a cool-core environment was studied with numerical simulations by \citet{vernaleo_problems_2006}. 
In that case, jets were shown to be incapable of producing heating/cooling balance, and were only able to delay the collapse of the core under radiative cooling by about $100$~Myr. The authors point out that the main difficulty is the formation of low-density channels through which the jet energy escapes, concluding that additional geometry or physics is necessary.  We find the same in our run 0000.  Additionally, the agreement between our run 0000 and the results of the latter authors goes further: they also find that their multiple jet beams end up inflating a single, very large cavity (their Fig. 6); strong gas circulation within the bubble; internal shocks in the vicinity of the jet channel due to self-collimation; and the presence of ripples propagating sideways.


\arif{Apart from the fixed-axis jets, \citet{vernaleo_problems_2006} also explored a scenario where the jets propagate through an ICM atmosphere that is rotating about the jet axis.  While this run has no physical analogue among our set of runs, the resulting configuration of the hot bubbles resembles our run 0030, in that the various formed cavities are part of an interconnected cone-like structure.  The authors found that while rotation enhanced the jet/ICM coupling, especially at the beginning of the simulation, the overall heating efficiency is comparable to that of our run 0030, including a fairly high accretion rate onto the core.}

\arif{In more recent simulations, \citet{yang_how_2016} investigated the effects of jets precessing around a fixed axis. In that case, the jets define a cone of influence, inside which most of the heating is located. The external region, though heated by thermalization of \emph{gentle circulation} driven by the jet cones, still features a reduced cooling flow. A more quantitative comparison will be possible after Paper II; however, we do note that qualitatively the \citet{yang_how_2016} precessing model is similar to our run 0030 in that the spatial  distribution of bubbles in the precessing jet model shows much more tightly distributed bubbles within a conical region about the precession axis, and the jets manage to establish a degree of dynamical balance between inflow and outflow even though weak transverse inflow persists. 
Additional qualitative similarities with our run 0030 
include that the youngest bubbles and a few bow-shocks/cocoons are always neatly distinguishable in the X-ray projection, and shocks are seen both inside the cocoons and propagating sideways in the ICM.}

\arif{
By and large, the bulk of recent jet simulations 
(e.g. \citealp{prasad_cool_2015,prasad_cool_2017,Prasad_turbulence_clouds_2018,yang_how_2016,cielo_backflow_2017,Li_AGN_2017,Gaspari_gaskinematics_2018})
have tended to focus on the thermal stability of the ICM, or the way gas accretes onto the black holes, or jets as a positive feedback phenomenon \citep{gaibler_jet-induced_2012}. Our simulations focus instead on the jet/ICM interaction \emph{per se} and therefore, a direct comparison is not possible.   However, we do note that even in those runs, the jets have been shown to excite significant turbulence and circulation flows and that these, in turn, can have a pronounced impact on the gas accretion rate (c.f. \citealp{prasad_cool_2017,Prasad_turbulence_clouds_2018,cielo_backflow_2017,cielo_compared_2018}).
}

\subsection{Considerations on re-orientation time} \label{sub:orientationtime}
The evolution of the jet-induced structure and its energetics that we presented descend also from the relatively short re-orientation time $\Delta t_{off}$ that we have chosen. We set $2$ out of $40$~Myr, corresponding to a jet duty cycle of $40/42 \sim 95\%$, comparable to that deduced by \cite{babul_isotropic_2013} based on their analysis of bubbles and jets in Perseus, M87 and CL0910. Alternative determination of the duty cycle by \cite{birzan_dutycycle_2012} shows that it may even approach $100\%$; however, values as low as $\sim 70\%$ have also appeared in the literature~\citep{gitti_skacoolcore_2015}.  A lower limit of $70\%$ would imply a re-orientation time of about $17$~Myr (since $40/57\sim70\%$); it is therefore important to consider the implications of a different re-orientation time.  

The consequence of a shorter re-orientation time, that is, of a virtually instantaneous re-orientation, are quite easily discussed. No difference is expected in cases of large re-orientation angles as in 2030 and 0090, as the jets would always propagate in a new direction. \arif{In the case of 0030, we also do not expect to see much difference: jets are still expected to flow into previously carved channels and bubbles whenever given the opportunity; as they change direction, they should excite ripples much like they do in the presented run.   We do, however, expect significant differences in the 0000 run.  In this case, the ripples are excited while a jet traverses an already open but slowly narrowing channel.  If $\Delta t_{off}$ is short, the channels will have less time to narrow and consequently, those localised shocks and ripples will be substantially reduced}\arifII{, further reducing the heating of the cluster core.}

A longer re-orientation time deserves a more exhaustive discussion, as we need to distinguish two cases, \arif{depending on whether the jets are off during re-orientation, as we have assumed, or not. Restricting ourselves to discussing whether successive jets will interact with previously carved channels and cavities, our tests suggest that the outcome depends on both the magnitude of $\Delta t_{off}$ and the angular displacement between successive jet pairs.}

\arif{
Let us first consider the ``long quiescence'' scenario where $\Delta t_{off}$ is a non-negligible fraction of the duration of a jet cycle, but still much shorter than the cooling time of the ICM in the cluster core. In the case of fixed-axis jets (run 0000), the shocks and ripples 
will become more prominent, since gas can have some time to cool and clump, obstructing the open pathways.  The same is true of turbulence and gas motions excited by the jets.
The cavities could then have time to detach from the jet chimney, which will be visible only for a fraction of the time. The new jets will however always impinge on the old bubbles, likely resulting in composite morphology similar to that seen in our run 0030, with individual cavities hosting a mixture of plasmas of different ages.  In the case of runs 2030 and 0090,  we do not expect to see any qualitative differences in terms of jet flows.   The case 0030 is more interesting, however.  If $\Delta t_{off}$ is sufficiently long that the bubbles have time to detach and rise away from the main trunk, the following jet will need to inflate a new bubble, resulting in a higher interaction efficiency between the jet and ICM in the cluster core.  We also note that increasing $\Delta t_{off}$,  the jet power would need to increase as well in order to maintain the same average power over time.  Cooling is a non-linear process, however, and one can imagine the ICM becoming thermally unstable in a catastrophic fashion during the off phase \citep{prasad_cool_2017}.  This deficiency could be mitigated in a self-regulating model where the jet activity is coupled to mass accretion onto the black hole.}

\arif{
The case of a longer re-orientation time, but where the jet remains active as it changes direction, could give rise to unique features.  For one, we would expect that as the jet changes direction -- and this is true for tilting jets as well as precessing jets -- one would expect to see transverse jet trails of the kind seen in the large-scale radio images of M87.   However, we already see features like this in our current runs:  they arise, as we have noted, when the jets encounter pre-existing cavities and change direction to expand into this cavity.  In cases where the angular change between successive jets is modest, we would expect the overall outcome to resemble the interconnected structure of current run 0030 (as in the similarities we have noted between the models of \citet{yang_how_2016} and our run 0030).  If successive jets can be offset from each other by a large angle (as in our runs 2030 and 0090), it is difficult to guess whether large tangential trails traced by the titling jets will create long-range channels between the bubbles or whether the jets will just proceed to inflate new bubbles.  These two cases have very different implications for the jet/ICM coupling.   
}

Overall, the re-orientation time $\Delta t_{off}$ appears to be as fundamental as the re-orientation angle. A degeneracy between the two is present, especially affecting the morphology (one main axis or not) and characteristics of the sources (ripples or not, isolated or multiple cavities) and the measurable angular distance between cavities. Yet no completely new feature is likely to emerge by varying $\Delta t_{off}$ alone, leading to the conclusion that the source panorama we present in the current section is complete.


\section{Summary and Conclusions}\label{sec:conclusion}

We have tested four models for AGN feedback from re-orienting jets in a cool-core cluster using numerical simulations performed with the FLASH code.
We have set up a spherical gaseous halo profile, initially in hydrostatic equilibrium within a $4.2\times10^{14}$~M$_\odot$ dark matter halo, and presenting a cool-cored entropy profile with an initial central cooling time of $150$~Myr.   We then shoot collimated jets from the halo centre, with a total kinetic power of $10^{45}$~erg/s, approximately equal to the initial radiative energy losses of the halo, and an internal Mach number of 3.
We have six/seven jet events per simulation, each lasting $40$~Myr and $2$~Myr between episodes ($\sim95\%$ duty cycle). The direction of each new jet is taken at random, but in each model we limit its angular displacement with respect to the previous jet axis in a different range: we constrain it to be always 0 (i.e. constant direction) in run 0000, or in the ranges $0$-$30$, $20$-$30$ and $0$-$90$ degrees, in our 0030, 2030 and 0090 models.

We visually compare the run by producing synthetic observations in soft and hard X-ray, identifying all the produced features (radio-lobes, cavities in the X-ray gas, bow-shocks, ripples and sound waves). We then compare the volume and energy of the cavities, as well as the mechanical expansion work performed against the ICM and the total energy available for heating. 
We also comment on how the different jet re-orientation models affect the gaseous halo stability by showing inflow/outflow maps and calculating the gas' total mass and energy within $50$, $100,$ and $150$~kpc.\\
\\
\noindent
Our main findings are summarised below.

\subsection{Cavity appearance and X-ray images}

\indent\par{\bf Each jet produces a cone of influence.} In agreement with previous numerical studies, jets/bubbles subtend a roughly conical region from the centre, opening to about $15$~degrees. A re-orientation angle larger than that will always result in detached rather than interconnected cavities, though some degeneracy between re-orientation angle and re-orientation time in this respect is more than likely. Modest re-orientation angles give rise to ripple-like features in the ICM.

\par{\bf The age of the plasma helps constrain the re-orientation history.} 
  Plasma age measurements are trustworthy indicators of the cavity ages, often more reliable than size or distance of the cavity from the centre, which may be affected by projection uncertainties. This is especially useful in the case of composite bubbles, where plasmas of different ages (i.e. coming from different jets) are seen to coexist.
  \par {\bf X-ray images may not show the whole cavity.} In the selected soft X-ray band ($[0.5,\,7]$~keV), all our synthetic images show only the highest-contrast region of the bubbles. This region coincides with the highest-temperature gas, directly evolved from the first-formed lobes. However, when viewed in pressure, it is clear that the cocoon material surrounding this hot region is part of the bubbles as well, and the actual bubbles may be much larger than what is seen in the X-ray.

\par {\bf X-ray images are not informative of the bubble's velocity structure.} Most of our bubbles have an inner vortex-ring-like velocity structure (as expected from light, supersonic jets), which affects their energetics and evolution. This is not evident from the soft X-ray images, which show only their quasi-spherical \emph{silhouettes}. This may lead to an underestimate of the bubbles kinetic energy, and has to be accounted for in any cavity model. 
  
\par {\bf The most realistic cluster features are in run 2030.} When producing synthetic soft X-ray images of the cluster, run 2030 shows alignment of several cavities filled with radio-emitting plasma, similar to what is observed in the X-ray gas around the jet of M87. The size, shape, and brightness of the latest bow-shock regions constitute a good match for the observations in the Perseus cluster. The gas pressure maps show good agreement as well, featuring mainly a young bow-shock region in a layered halo, with absence of extra shock-driven features or large bubbles, very prominent in the other runs. 

\arifII{
\par {\bf Exceptionally large observed cavities are likely due to multiple jets feeding the same bubble.}   The unusually large cavities in Hydra A and MS0735.6+7421 are typically interpreted as products of a single exceptionally powerful jet outburst.  In light of the close resemblance between these cavities and the cavities in our fixed-axis jet run (0000), an alternate explanation is that the cavities in Hydra and MS0735 have been fed and inflated by multiple episodes of otherwise typical jets fired in the same direction.
}
  
\subsection{Cavity energetics}
   \indent\par {\bf Smaller re-orientation angles imply larger cavities with higher kinetic energy}. For small re-orientation angles (run 0000 and 0030) jets inflate only one connected cavity structure, which extends out to large distances from the cluster centre.  This results in more efficient transfer of jet power to the cavities and allows for more efficient large-scale flows within the bubbles.  In run 0000, the cavity kinetic energy is as much as $\sim 50\%$ of the thermal energy.  This energy is be readily available as a heat source for the efficiently cooling ICM in the cluster core.  
  \par {\bf Re-orienting jets transfer more energy to the X-ray gas.} When considering the sum of all the cavities, the fixed-axis jets (i.e our run 0000) can transfer only up to about $65\%$ of their energy to the X-ray gas, versus $\sim 80\%$ for the re-orienting cases. 
This happens despite the jets in run 0000 showing larger cavity volumes and a larger value of the mechanical expansion work.  \arifII{On the other hand, this difference in global coupling efficiency does not represent the whole picture.  Where the jets inflate the cavities, and whether the cavities are independent detached structures or part of an extended interconnected structure, is just as, if not more, important (see Sect. \ref{conclude_coreheating} below).}

  \par {\bf The integrated cooling losses are reduced in the presence re-orienting jets.} The total cooling luminosity of the halo, in the absence of jets, increases dramatically in a runaway process.  In agreement with \citet{vernaleo_problems_2006}, we find that jets in a fixed direction can only delay this fate by $\sim100-200$~Myr.  Re-orientating jets, as in our runs 2030 and 0090, are highly effective at altering the cooling profile of the cluster core, restricting the increase in the cooling luminosity to about $5\%$ over $250$~Myrs.  In these models, heating-cooling balance is within reach with a slight fine-tuning of the jet profile (energy and momentum flux), which we have not attempted.   
  
\subsection{Cool-core heating and stability}\label{conclude_coreheating}
 
  \par {\bf Re-orienting jets heat the cluster gas core more efficiently.} When \sI{a young jet} \arifII{inflate\sI{s a} detached cavit\sI{y in a new direction, such a cavity} form\sI{s} within the cluster core. As a result, the core is subject to new shocks after every jet episode. \sI{Even at later times, when bubbles start to rise, there are always many more bubbles within the innermost $100$~kpc, whose distribution --- in addition --- spans a much larger solid angle, because the jet orientation is more isotropic. These combined effects} hinder formation of cooling flows and also disrupt coherent backflows that could promote runaway thermal instability within the core.}
  \par {\bf Re-orienting jets are more effective at establishing a dynamical inflow-outflow balance.} The interplay between \arifII{jet-induced outflows and inward-directed backflow-augmented cooling flows 
creates an alternating inflow-outflow pattern with distance from the cluster centre. In our moderately-to-strongly reorientating jet models, the outflows manage to significantly counteract the inflows, preventing most of the mass increase within the central $50$~kpc that we observe in all the other runs around $250$~Myrs. Specifically,\protect\LEt{The meaning of `in fin fact' is ambiguous here and, also throughout the paper; I have suggested an alternative but please feel free to suggest another if deemed to be more fitting.} the outflows drive the net expulsion of gas from within the central $100$ and $150$~kpc.  Similar features are present in the plot of the core gas temperature.}

  
\begin{acknowledgements}
  This work was granted access to the HPC resources of CINES under the allocations x2016046955 and A0020406955 made by GENCI. The work of S.C. has been supported by the ERC Project No. 267117 (DARK) hosted by Universit\'{e} Pierre et Marie Curie (UPMC) - Paris 6 and the ERC Project No. 614199 (BLACK) at  Centre National De La Recherche Scientifique (CNRS). AB acknowledges support from NSERC (Canada) through the Discovery Grant program, from Institut Lagrange de Paris (ILP), and the Pauli Center for Theoretical Studies ETH UZH. S.C. is also grateful to the University of Zurich's Institute for Computational Sciences's Center for Theoretical Astrophysics and Cosmology and the Institut d'Astrophysique de Paris for  hospitality during his recent sabbatical visit.  S.C. thanks Ewan O'Sullivan for the precious advise on the production of realistic X-ray images.
\end{acknowledgements}

\bibliographystyle{aa}
\bibliography{all}

\begin{thebibliography}{123}
\expandafter\ifx\csname natexlab\endcsname\relax\def\natexlab#1{#1}\fi

\bibitem[{Antonuccio-Delogu \& Silk(2008)}]{antonuccio-delogu_active_2008}
Antonuccio-Delogu, V. \& Silk, J. 2008, \mnras, 389, 1750

\bibitem[{{Antonuccio-Delogu} \& {Silk}(2010)}]{antonuccio-delogu_feeding_2010}
{Antonuccio-Delogu}, V. \& {Silk}, J. 2010, in Astronomical Society of the
  Pacific Conference Series, Vol. 427, Accretion and Ejection in AGN: a Global
  View, ed. L.~{Maraschi}, G.~{Ghisellini}, R.~{Della Ceca}, \& F.~{Tavecchio},
  343

\bibitem[{Babul {et~al.}(2002)Babul, Balogh, Lewis, \&
  Poole}]{babul_physical_2002}
Babul, A., Balogh, M.~L., Lewis, G.~F., \& Poole, G.~B. 2002, Monthly Notices
  of the Royal Astronomical Society, 330, 329

\bibitem[{Babul {et~al.}(2013)Babul, Sharma, \&
  Reynolds}]{babul_isotropic_2013}
Babul, A., Sharma, P., \& Reynolds, C.~S. 2013, \apj, 768, 11

\bibitem[{{Babyk} {et~al.}(2018){Babyk}, {McNamara}, {Nulsen}, {Russell},
  {Vantyghem}, {Hogan}, \& {Pulido}}]{Babyk_entropyprofile_2018}
{Babyk}, I.~V., {McNamara}, B.~R., {Nulsen}, P.~E.~J., {et~al.} 2018, ArXiv
  e-prints [\eprint[arXiv]{1802.02589}]

\bibitem[{Biffi {et~al.}(2013)Biffi, Dolag, \&
  B{\~A}{\^A}¶hringer}]{biffi_investigating_2013}
Biffi, V., Dolag, K., \& B{\~A}{\^A}¶hringer, H. 2013, \mnras, 428, 1395

\bibitem[{Binney \& Tabor(1995)}]{binney_evolving_1995}
Binney, J. \& Tabor, G. 1995, Monthly Notices of the Royal Astronomical
  Society, 276, 663

\bibitem[{{B{\^i}rzan} {et~al.}(2012){B{\^i}rzan}, {Rafferty}, {Nulsen},
  {McNamara}, {R{\"o}ttgering}, {Wise}, \& {Mittal}}]{birzan_dutycycle_2012}
{B{\^i}rzan}, L., {Rafferty}, D.~A., {Nulsen}, P.~E.~J., {et~al.} 2012, \mnras,
  427, 3468

\bibitem[{{Boselli} {et~al.}(2016){Boselli}, {Cuillandre}, {Fossati},
  {Boissier}, {Bomans}, {Consolandi}, {Anselmi}, {Cortese}, {C{\^o}t{\'e}},
  {Durrell}, {Ferrarese}, {Fumagalli}, {Gavazzi}, {Gwyn}, {Hensler}, {Sun}, \&
  {Toloba}}]{boselli_ionized_2017}
{Boselli}, A., {Cuillandre}, J.~C., {Fossati}, M., {et~al.} 2016, \aap, 587,
  A68

\bibitem[{Brennen(1995)}]{brennen_bubble_1995}
Brennen, C.~E. 1995, Cavitation Bubble Dynamics and Noise Production

\bibitem[{{Br{\"u}ggen}(2003)}]{bruggen_bubbles_2003}
{Br{\"u}ggen}, M. 2003, \apj, 592, 839

\bibitem[{Brüggen(2003)}]{Bruggen2003}
Brüggen, M. 2003, The Astrophysical Journal, 592, 839

\bibitem[{{Campanelli} {et~al.}(2007){Campanelli}, {Lousto}, {Zlochower}, \&
  {Merritt}}]{Campanelli_SpinFlip_2007}
{Campanelli}, M., {Lousto}, C., {Zlochower}, Y., \& {Merritt}, D. 2007, \apjl,
  659, L5

\bibitem[{{Cavagnolo} {et~al.}(2008){Cavagnolo}, {Donahue}, {Voit}, \&
  {Sun}}]{Cavagnolo2008}
{Cavagnolo}, K.~W., {Donahue}, M., {Voit}, G.~M., \& {Sun}, M. 2008, \apjl,
  683, L107

\bibitem[{Cavagnolo {et~al.}(2009)Cavagnolo, Donahue, Voit, \&
  Sun}]{cavagnolo_intracluster_2009}
Cavagnolo, K.~W., Donahue, M., Voit, G.~M., \& Sun, M. 2009, \apjs, 182, 12

\bibitem[{Churazov {et~al.}(2016)Churazov, Arevalo, Forman, Jones,
  Schekochihin, Vikhlinin, \& Zhuravleva}]{churazov_arithmetic_2016}
Churazov, E., Arevalo, P., Forman, W., {et~al.} 2016, \mnras, 463, 1057

\bibitem[{{Churazov} {et~al.}(2002){Churazov}, {Sunyaev}, {Forman}, \&
  {Boehringer}}]{churazov_calorimeter_2002}
{Churazov}, E., {Sunyaev}, R., {Forman}, W., \& {Boehringer}, H. 2002, \mnras,
  332, 729

\bibitem[{Cielo {et~al.}(2014)Cielo, Antonuccio-Delogu, Macci\`{o}, Romeo, \&
  Silk}]{cielo_3d_2014}
Cielo, S., Antonuccio-Delogu, V., Macci\`{o}, A.~V., Romeo, A.~D., \& Silk, J.
  2014, \mnras, 439, 2903

\bibitem[{{Cielo} {et~al.}(2017){Cielo}, {Antonuccio-Delogu}, {Silk}, \&
  {Romeo}}]{cielo_backflow_2017}
{Cielo}, S., {Antonuccio-Delogu}, V., {Silk}, J., \& {Romeo}, A.~D. 2017,
  \mnras, 467, 4526

\bibitem[{{Cielo} {et~al.}(2018){Cielo}, {Bieri}, {Volonteri}, {Wagner}, \&
  {Dubois}}]{cielo_compared_2018}
{Cielo}, S., {Bieri}, R., {Volonteri}, M., {Wagner}, A.~Y., \& {Dubois}, Y.
  2018, \mnras [\eprint[arXiv]{1712.03955}]

\bibitem[{Ciotti \& Ostriker(2001)}]{ciotti_cooling_2001}
Ciotti, L. \& Ostriker, J.~P. 2001, The Astrophysical Journal, 551, 131

\bibitem[{{Clautice} {et~al.}(2016){Clautice}, {Perlman}, {Georganopoulos},
  {Lister}, {Tombesi}, {Cara}, {Marshall}, {Hogan}, \&
  {Kazanas}}]{clautice_spectacular_2016}
{Clautice}, D., {Perlman}, E.~S., {Georganopoulos}, M., {et~al.} 2016, \apj,
  826, 109

\bibitem[{{Cowie} {et~al.}(1980){Cowie}, {Fabian}, \& {Nulsen}}]{Cowie1980}
{Cowie}, L.~L., {Fabian}, A.~C., \& {Nulsen}, P.~E.~J. 1980, \mnras, 191, 399

\bibitem[{{Crawford} {et~al.}(1999){Crawford}, {Allen}, {Ebeling}, {Edge}, \&
  {Fabian}}]{Crawford1999}
{Crawford}, C.~S., {Allen}, S.~W., {Ebeling}, H., {Edge}, A.~C., \& {Fabian},
  A.~C. 1999, \mnras, 306, 857

\bibitem[{{de Gouveia Dal Pino}(1999)}]{DeflectedJets_1999}
{de Gouveia Dal Pino}, E.~M. 1999, \apj, 526, 862

\bibitem[{{Doria} {et~al.}(2012){Doria}, {Gitti}, {Ettori}, {Brighenti},
  {Nulsen}, \& {McNamara}}]{Doria_RBS797_2012}
{Doria}, A., {Gitti}, M., {Ettori}, S., {et~al.} 2012, \apj, 753, 47

\bibitem[{{Duan} \& {Guo}(2018)}]{DuanGuo2018}
{Duan}, X. \& {Guo}, F. 2018, ArXiv e-prints [\eprint[arXiv]{1805.02689}]

\bibitem[{{Dunn} {et~al.}(2006){Dunn}, {Fabian}, \&
  {Sanders}}]{Dunn_BHPrecess_2006}
{Dunn}, R.~J.~H., {Fabian}, A.~C., \& {Sanders}, J.~S. 2006, \mnras, 366, 758

\bibitem[{{Fabian}(2012)}]{fabian_observational_2012}
{Fabian}, A.~C. 2012, \araa, 50, 455

\bibitem[{{Fabian} {et~al.}(2011){Fabian}, {Sanders}, {Allen}, {Canning},
  {Churazov}, {Crawford}, {Forman}, {Gabany}, {Hlavacek-Larrondo}, {Johnstone},
  {Russell}, {Reynolds}, {Salom{\'e}}, {Taylor}, \&
  {Young}}]{fabian_perseus_2011}
{Fabian}, A.~C., {Sanders}, J.~S., {Allen}, S.~W., {et~al.} 2011, \mnras, 418,
  2154

\bibitem[{{Fabian} {et~al.}(2000){Fabian}, {Sanders}, {Ettori}, {Taylor},
  {Allen}, {Crawford}, {Iwasawa}, {Johnstone}, \& {Ogle}}]{fabian_perseus_2000}
{Fabian}, A.~C., {Sanders}, J.~S., {Ettori}, S., {et~al.} 2000, \mnras, 318,
  L65

\bibitem[{{Fabian} {et~al.}(2017){Fabian}, {Walker}, {Russell}, {Pinto},
  {Sanders}, \& {Reynolds}}]{fabian_transport_2017}
{Fabian}, A.~C., {Walker}, S.~A., {Russell}, H.~R., {et~al.} 2017, \mnras, 464,
  L1

\bibitem[{{Falceta-Gon{\c c}alves} {et~al.}(2010){Falceta-Gon{\c c}alves},
  {Caproni}, {Abraham}, {Teixeira}, \& {de Gouveia Dal
  Pino}}]{Precess_Jets_2010}
{Falceta-Gon{\c c}alves}, D., {Caproni}, A., {Abraham}, Z., {Teixeira}, D.~M.,
  \& {de Gouveia Dal Pino}, E.~M. 2010, \apjl, 713, L74

\bibitem[{{Forman} {et~al.}(2017){Forman}, {Churazov}, {Jones}, {Heinz},
  {Kraft}, \& {Vikhlinin}}]{forman_M87_2017}
{Forman}, W., {Churazov}, E., {Jones}, C., {et~al.} 2017, \apj, 844, 122

\bibitem[{{Forman} {et~al.}(2005){Forman}, {Nulsen}, {Heinz}, {Owen}, {Eilek},
  {Vikhlinin}, {Markevitch}, {Kraft}, {Churazov}, \& {Jones}}]{forman_M87_2005}
{Forman}, W., {Nulsen}, P., {Heinz}, S., {et~al.} 2005, \apj, 635, 894

\bibitem[{{Franchini} {et~al.}(2016){Franchini}, {Lodato}, \&
  {Facchini}}]{Franchini_BHprecess_2016}
{Franchini}, A., {Lodato}, G., \& {Facchini}, S. 2016, \mnras, 455, 1946

\bibitem[{Fryxell {et~al.}(2000)Fryxell, Olson, Ricker, Timmes, Zingale, Lamb,
  MacNeice, Rosner, Truran, \& Tufo}]{fryxell_flash:_2000}
Fryxell, B., Olson, K., Ricker, P., {et~al.} 2000, Astrophys. J. Supp., 131,
  273

\bibitem[{Gaibler {et~al.}(2010)Gaibler, Khochfar, \&
  Krause}]{gaibler_asymmetries_2010}
Gaibler, V., Khochfar, S., \& Krause, M. 2010

\bibitem[{Gaibler {et~al.}(2012)Gaibler, Khochfar, Krause, \&
  Silk}]{gaibler_jet-induced_2012}
Gaibler, V., Khochfar, S., Krause, M., \& Silk, J. 2012, \mnras, 425, 438

\bibitem[{{Gaspari} {et~al.}(2018){Gaspari}, {McDonald}, {Hamer}, {Brighenti},
  {Temi}, {Gendron-Marsolais}, {Hlavacek-Larrondo}, {Edge}, {Werner}, {Tozzi},
  {Sun}, {Stone}, {Tremblay}, {Hogan}, {Eckert}, {Ettori}, {Yu}, {Biffi}, \&
  {Planelles}}]{Gaspari_gaskinematics_2018}
{Gaspari}, M., {McDonald}, M., {Hamer}, S.~L., {et~al.} 2018, \apj, 854, 167

\bibitem[{{Gaspari} {et~al.}(2013){Gaspari}, {Ruszkowski}, \&
  {Oh}}]{gaspari_cca_2013}
{Gaspari}, M., {Ruszkowski}, M., \& {Oh}, S.~P. 2013, \mnras, 432, 3401

\bibitem[{{Gaspari} {et~al.}(2017){Gaspari}, {Temi}, \&
  {Brighenti}}]{Gaspari2017}
{Gaspari}, M., {Temi}, P., \& {Brighenti}, F. 2017, \mnras, 466, 677

\bibitem[{{Gerosa} {et~al.}(2015){Gerosa}, {Kesden}, {O'Shaughnessy}, {Klein},
  {Berti}, {Sperhake}, \& {Trifir{\`o}}}]{Gerosa_BHPrecession_2015}
{Gerosa}, D., {Kesden}, M., {O'Shaughnessy}, R., {et~al.} 2015, Physical Review
  Letters, 115, 141102

\bibitem[{{Gitti} {et~al.}(2006{\natexlab{a}}){Gitti}, {Feretti}, \&
  {Schindler}}]{Gitti_PrecessingJet_2006}
{Gitti}, M., {Feretti}, L., \& {Schindler}, S. 2006{\natexlab{a}}, \aap, 448,
  853

\bibitem[{{Gitti} {et~al.}(2006{\natexlab{b}}){Gitti}, {Feretti}, \&
  {Schindler}}]{Gitti_RBS797_2006}
{Gitti}, M., {Feretti}, L., \& {Schindler}, S. 2006{\natexlab{b}}, \aap, 448,
  853

\bibitem[{{Gitti} {et~al.}(2015){Gitti}, {Tozzi}, {Brunetti}, {Cassano},
  {Dallacasa}, {Edge}, {Ettori}, {Feretti}, {Ferrari}, {Giacintucci},
  {Giovannini}, {Hogan}, \& {Venturi}}]{gitti_skacoolcore_2015}
{Gitti}, M., {Tozzi}, P., {Brunetti}, G., {et~al.} 2015, Advancing Astrophysics
  with the Square Kilometre Array (AASKA14), 76

\bibitem[{Guo(2015)}]{guo_shape_2015}
Guo, F. 2015, \apj, 803, 48

\bibitem[{Guo(2016)}]{guo_importance_2016}
Guo, F. 2016, \apj, 826, 17

\bibitem[{{Hatch} {et~al.}(2007){Hatch}, {Crawford}, \& {Fabian}}]{Hatch2007}
{Hatch}, N.~A., {Crawford}, C.~S., \& {Fabian}, A.~C. 2007, \mnras, 380, 33

\bibitem[{{Hattori} {et~al.}(1995){Hattori}, {Yoshida}, \&
  {Habe}}]{Hattori1995}
{Hattori}, M., {Yoshida}, T., \& {Habe}, A. 1995, \mnras, 275, 1195

\bibitem[{{Heckman} {et~al.}(1989){Heckman}, {Baum}, {van Breugel}, \&
  {McCarthy}}]{Heckman1989}
{Heckman}, T.~M., {Baum}, S.~A., {van Breugel}, W.~J.~M., \& {McCarthy}, P.
  1989, \apj, 338, 48

\bibitem[{Heinz {et~al.}(2006)Heinz, Brueggen, Young, \&
  Levesque}]{heinz_answer_2006}
Heinz, S., Brueggen, M., Young, A., \& Levesque, E. 2006,
  Mon.Not.Roy.Astron.Soc.Lett., 373, L65

\bibitem[{Hlavacek-Larrondo {et~al.}(2012)Hlavacek-Larrondo, Fabian, Edge,
  Ebeling, Sanders, Hogan, \& Taylor}]{hlavacek-larrondo_extreme_2012}
Hlavacek-Larrondo, J., Fabian, A.~C., Edge, A.~C., {et~al.} 2012, \mnras, 421,
  1360

\bibitem[{{Hobbs} {et~al.}(2011){Hobbs}, {Nayakshin}, {Power}, \&
  {King}}]{Hobbs2011}
{Hobbs}, A., {Nayakshin}, S., {Power}, C., \& {King}, A. 2011, \mnras, 413,
  2633

\bibitem[{{Kesden} {et~al.}(2010){Kesden}, {Sperhake}, \&
  {Berti}}]{Kesden_SpinFlip_2010}
{Kesden}, M., {Sperhake}, U., \& {Berti}, E. 2010, \prd, 81, 084054

\bibitem[{Komatsu {et~al.}(2011)Komatsu, Smith, Dunkley, Bennett, \&
  {Gold}}]{komatsu_seven-year_2011}
Komatsu, E., Smith, K.~M., Dunkley, J., Bennett, C.~L., \& {Gold}. 2011, \apjs,
  192, 18

\bibitem[{{Lau} {et~al.}(2017){Lau}, {Gaspari}, {Nagai}, \&
  {Coppi}}]{lauetal_perseusar_2017}
{Lau}, E.~T., {Gaspari}, M., {Nagai}, D., \& {Coppi}, P. 2017, \apj, 849, 54

\bibitem[{{Li} \& {Bryan}(2014)}]{Li_TI_2014}
{Li}, Y. \& {Bryan}, G.~L. 2014, \apj, 789, 153

\bibitem[{{Li} {et~al.}(2017){Li}, {Ruszkowski}, \& {Bryan}}]{Li_AGN_2017}
{Li}, Y., {Ruszkowski}, M., \& {Bryan}, G.~L. 2017, \apj, 847, 106

\bibitem[{Liang {et~al.}(2016)Liang, Durier, Babul, Dav{\~A}{\^A}©,
  Oppenheimer, Katz, Fardal, \& Quinn}]{liang_growth_2016}
Liang, L., Durier, F., Babul, A., {et~al.} 2016, \mnras, 456, 4266

\bibitem[{{Lodato} \& {Pringle}(2006)}]{LodatoPringle_2006}
{Lodato}, G. \& {Pringle}, J.~E. 2006, \mnras, 368, 1196

\bibitem[{L{\"o}hner(1987)}]{lohner_adaptive_1987}
L{\"o}hner, R. 1987, Computer Methods in Applied Mechanics and Engineering, 61,
  323

\bibitem[{{Maller} \& {Bullock}(2004)}]{MallerBullock2004}
{Maller}, A.~H. \& {Bullock}, J.~S. 2004, \mnras, 355, 694

\bibitem[{McCarthy {et~al.}(2008)McCarthy, Babul, Bower, \&
  Balogh}]{mccarthy_towards_2008}
McCarthy, I.~G., Babul, A., Bower, R.~G., \& Balogh, M.~L. 2008, Monthly
  Notices of the Royal Astronomical Society, 386, 1309

\bibitem[{{McDonald} {et~al.}(2010){McDonald}, {Veilleux}, {Rupke}, \&
  {Mushotzky}}]{McDonald_Halpha_2010}
{McDonald}, M., {Veilleux}, S., {Rupke}, D.~S.~N., \& {Mushotzky}, R. 2010,
  \apj, 721, 1262

\bibitem[{{McNamara} {et~al.}(2009){McNamara}, {Kazemzadeh}, {Rafferty},
  {B{\^i}rzan}, {Nulsen}, {Kirkpatrick}, \&
  {Wise}}]{mcnamara_ultramassiveAGN_2009}
{McNamara}, B.~R., {Kazemzadeh}, F., {Rafferty}, D.~A., {et~al.} 2009, \apj,
  698, 594

\bibitem[{McNamara \& Nulsen(2007)}]{mcnamara_heating_2007}
McNamara, B.~R. \& Nulsen, P. E.~J. 2007, Annual Review of Astronomy and
  Astrophysics, 45, 117

\bibitem[{McNamara \& Nulsen(2012)}]{mcnamara_mechanical_2012}
McNamara, B.~R. \& Nulsen, P. E.~J. 2012, New Journal of Physics, 14, 055023

\bibitem[{{McNamara} {et~al.}(2005){McNamara}, {Nulsen}, {Wise}, {Rafferty},
  {Carilli}, {Sarazin}, \& {Blanton}}]{mcnamara_heatingnature_2005}
{McNamara}, B.~R., {Nulsen}, P.~E.~J., {Wise}, M.~W., {et~al.} 2005, \nat, 433,
  45

\bibitem[{{McNamara} {et~al.}(2001){McNamara}, {Wise}, {Nulsen}, {David},
  {Carilli}, {Sarazin}, {O'Dea}, {Houck}, {Donahue}, {Baum}, {Voit},
  {O'Connell}, \& {Koekemoer}}]{mcnamara_a2597_2001}
{McNamara}, B.~R., {Wise}, M.~W., {Nulsen}, P.~E.~J., {et~al.} 2001, \apjl,
  562, L149

\bibitem[{{Mendoza} \& {Longair}(2001)}]{Mendoza_2001}
{Mendoza}, S. \& {Longair}, M.~S. 2001, \mnras, 324, 149

\bibitem[{{Mendygral} {et~al.}(2012){Mendygral}, {Jones}, \&
  {Dolag}}]{mendygral_mhdjets_2012}
{Mendygral}, P.~J., {Jones}, T.~W., \& {Dolag}, K. 2012, \apj, 750, 166

\bibitem[{{Mendygral} {et~al.}(2011){Mendygral}, {O'Neill}, \&
  {Jones}}]{mendygral_syntheticAGN_2011}
{Mendygral}, P.~J., {O'Neill}, S.~M., \& {Jones}, T.~W. 2011, \apj, 730, 100

\bibitem[{{Merritt} \& {Ekers}(2002)}]{Merritt_SpinFlip_2002}
{Merritt}, D. \& {Ekers}, R.~D. 2002, Science, 297, 1310

\bibitem[{{Merritt} \& {Vasiliev}(2012)}]{Merritt_BHPrecess_2012}
{Merritt}, D. \& {Vasiliev}, E. 2012, \prd, 86, 102002

\bibitem[{{Morsony} {et~al.}(2010){Morsony}, {Heinz}, {Br{\"u}ggen}, \&
  {Ruszkowski}}]{morsony_swimming_2010}
{Morsony}, B.~J., {Heinz}, S., {Br{\"u}ggen}, M., \& {Ruszkowski}, M. 2010,
  \mnras, 407, 1277

\bibitem[{{Nawaz} {et~al.}(2016){Nawaz}, {Bicknell}, {Wagner}, {Sutherland}, \&
  {McNamara}}]{Nawaz_JetPrecess_2016}
{Nawaz}, M.~A., {Bicknell}, G.~V., {Wagner}, A.~Y., {Sutherland}, R.~S., \&
  {McNamara}, B.~R. 2016, \mnras, 458, 802

\bibitem[{Neumayer {et~al.}(2007)Neumayer, Cappellari, Reunanen, Rix, van~der
  Werf, de~Zeeuw, \& Davies}]{neumayer_central_2007}
Neumayer, N., Cappellari, M., Reunanen, J., {et~al.} 2007, \apj, 671, 1329

\bibitem[{Newman {et~al.}(2013)Newman, Treu, Ellis, Sand, Nipoti, Richard, \&
  Jullo}]{newman_density_2013}
Newman, A.~B., Treu, T., Ellis, R.~S., {et~al.} 2013, \apj, 765, 24

\bibitem[{{Nipoti} \& {Binney}(2004)}]{NipotiBinney2004}
{Nipoti}, C. \& {Binney}, J. 2004, \mnras, 349, 1509

\bibitem[{{Nulsen} {et~al.}(2005){Nulsen}, {McNamara}, {Wise}, \&
  {David}}]{nulsen_hydraA_2005}
{Nulsen}, P.~E.~J., {McNamara}, B.~R., {Wise}, M.~W., \& {David}, L.~P. 2005,
  \apj, 628, 629

\bibitem[{{Nusser} {et~al.}(2006){Nusser}, {Silk}, \&
  {Babul}}]{nusser_suppressing_2006}
{Nusser}, A., {Silk}, J., \& {Babul}, A. 2006, \mnras, 373, 739

\bibitem[{{Omma} \& {Binney}(2004)}]{Omma_2004}
{Omma}, H. \& {Binney}, J. 2004, \mnras, 350, L13

\bibitem[{{Omma} {et~al.}(2004){Omma}, {Binney}, {Bryan}, \&
  {Slyz}}]{Omma_Binney_Bryan_Slyz_2004}
{Omma}, H., {Binney}, J., {Bryan}, G., \& {Slyz}, A. 2004, \mnras, 348, 1105

\bibitem[{O'Neill \& Jones(2010)}]{oneill_intermittentjets_2010}
O'Neill, S.~M. \& Jones, T.~W. 2010, The Astrophysical Journal, 710, 180

\bibitem[{O'Sullivan {et~al.}(2012)O'Sullivan, Giacintucci, Babul,
  Raychaudhury, Venturi, Bildfell, Mahdavi, Oonk, Murray, Hoekstra, \&
  Donahue}]{osullivan_giant_2012}
O'Sullivan, E., Giacintucci, S., Babul, A., {et~al.} 2012, Monthly Notices of
  the Royal Astronomical Society, 424, 2971

\bibitem[{{O'Sullivan} {et~al.}(2017){O'Sullivan}, {Ponman}, {Kolokythas},
  {Raychaudhury}, {Babul}, {Vrtilek}, {David}, {Giacintucci}, {Gitti}, \&
  {Haines}}]{CLoGS_GroupsEntropyProfile_2017}
{O'Sullivan}, E., {Ponman}, T.~J., {Kolokythas}, K., {et~al.} 2017, \mnras,
  472, 1482

\bibitem[{{Owen} {et~al.}(2000){Owen}, {Eilek}, \&
  {Kassim}}]{owen_M87radio_2000}
{Owen}, F.~N., {Eilek}, J.~A., \& {Kassim}, N.~E. 2000, \apj, 543, 611

\bibitem[{{Panagoulia} {et~al.}(2014){Panagoulia}, {Fabian}, \&
  {Sanders}}]{Panagoulia_CoreEntropyProfile_2014}
{Panagoulia}, E.~K., {Fabian}, A.~C., \& {Sanders}, J.~S. 2014, \mnras, 438,
  2341

\bibitem[{Perucho {et~al.}(2014)Perucho, Mart{\'i}, Laing, \&
  Hardee}]{perucho_deceleration_2014}
Perucho, M., Mart{\'i}, J.~M., Laing, R.~A., \& Hardee, P.~E. 2014, \mnras,
  441, 1488

\bibitem[{{Pizzolato} \& {Soker}(2005{\natexlab{a}})}]{Pizzolato_BBH_2005}
{Pizzolato}, F. \& {Soker}, N. 2005{\natexlab{a}}, Advances in Space Research,
  36, 762

\bibitem[{{Pizzolato} \&
  {Soker}(2005{\natexlab{b}})}]{Pizzolato_AGNClouds_2005}
{Pizzolato}, F. \& {Soker}, N. 2005{\natexlab{b}}, \apj, 632, 821

\bibitem[{{Pizzolato} \& {Soker}(2010)}]{Pizzolato_AGNClouds_2010}
{Pizzolato}, F. \& {Soker}, N. 2010, \mnras, 408, 961

\bibitem[{{Poole} {et~al.}(2006){Poole}, {Fardal}, {Babul}, {McCarthy},
  {Quinn}, \& {Wadsley}}]{poole_mergers_2006}
{Poole}, G.~B., {Fardal}, M.~A., {Babul}, A., {et~al.} 2006, \mnras, 373, 881

\bibitem[{{Pope} {et~al.}(2010){Pope}, {Babul}, {Pavlovski}, {Bower}, \&
  {Dotter}}]{pope_transport_2010}
{Pope}, E.~C.~D., {Babul}, A., {Pavlovski}, G., {Bower}, R.~G., \& {Dotter}, A.
  2010, \mnras, 406, 2023

\bibitem[{Prasad {et~al.}(2015)Prasad, Sharma, \& Babul}]{prasad_cool_2015}
Prasad, D., Sharma, P., \& Babul, A. 2015, \apj, 811, 108

\bibitem[{Prasad {et~al.}(2017)Prasad, Sharma, \& Babul}]{prasad_cool_2017}
Prasad, D., Sharma, P., \& Babul, A. 2017, \mnras

\bibitem[{{Prasad} {et~al.}(2018){Prasad}, {Sharma}, \&
  {Babul}}]{Prasad_turbulence_clouds_2018}
{Prasad}, D., {Sharma}, P., \& {Babul}, A. 2018, ArXiv e-prints
  [\eprint[arXiv]{1801.04282}]

\bibitem[{{Randall} {et~al.}(2015){Randall}, {Nulsen}, {Jones}, {Forman},
  {Bulbul}, {Clarke}, {Kraft}, {Blanton}, {David}, {Werner}, {Sun}, {Donahue},
  {Giacintucci}, \& {Simionescu}}]{randall_ngc5813_2015}
{Randall}, S.~W., {Nulsen}, P.~E.~J., {Jones}, C., {et~al.} 2015, \apj, 805,
  112

\bibitem[{Rephaeli \& Silk(1995)}]{rephaeli_energetic_1995}
Rephaeli, Y. \& Silk, J. 1995, The Astrophysical Journal, 442, 91

\bibitem[{{Revaz} {et~al.}(2008){Revaz}, {Combes}, \& {Salom{\'e}}}]{Revaz2008}
{Revaz}, Y., {Combes}, F., \& {Salom{\'e}}, P. 2008, \aap, 477, L33

\bibitem[{{Reynolds} {et~al.}(2001){Reynolds}, {Heinz}, \&
  {Begelman}}]{Reynolds_2001}
{Reynolds}, C.~S., {Heinz}, S., \& {Begelman}, M.~C. 2001, \apjl, 549, L179

\bibitem[{{Reynolds} {et~al.}(2002){Reynolds}, {Heinz}, \&
  {Begelman}}]{Reynolds_2002}
{Reynolds}, C.~S., {Heinz}, S., \& {Begelman}, M.~C. 2002, \mnras, 332, 271

\bibitem[{Roberts {et~al.}(2015)Roberts, Cohen, Lu, Saripalli, \&
  Subrahmanyan}]{roberts_abundance_2015}
Roberts, D.~H., Cohen, J.~P., Lu, J., Saripalli, L., \& Subrahmanyan, R. 2015,
  \apjs, 220, 7

\bibitem[{{Saxton} {et~al.}(2005){Saxton}, {Bicknell}, {Sutherland}, \&
  {Midgley}}]{Saxton_2005}
{Saxton}, C.~J., {Bicknell}, G.~V., {Sutherland}, R.~S., \& {Midgley}, S. 2005,
  \mnras, 359, 781

\bibitem[{{Saxton} {et~al.}(2001){Saxton}, {Sutherland}, \&
  {Bicknell}}]{Saxton2001}
{Saxton}, C.~J., {Sutherland}, R.~S., \& {Bicknell}, G.~V. 2001, \apj, 563, 103

\bibitem[{{Sharma} {et~al.}(2012){Sharma}, {McCourt}, {Quataert}, \&
  {Parrish}}]{Sharma_TI_2012}
{Sharma}, P., {McCourt}, M., {Quataert}, E., \& {Parrish}, I.~J. 2012, \mnras,
  420, 3174

\bibitem[{{Simionescu} {et~al.}(2017){Simionescu}, {Werner}, {Mantz}, {Allen},
  \& {Urban}}]{simionescu2017virgo}
{Simionescu}, A., {Werner}, N., {Mantz}, A., {Allen}, S.~W., \& {Urban}, O.
  2017, \mnras, 469, 1476

\bibitem[{Soker(2016)}]{soker_jet_2016}
Soker, N. 2016, New Astronomy Reviews, 75, 1

\bibitem[{{Soker} \& {Bisker}(2006)}]{soker_jetbending_2006}
{Soker}, N. \& {Bisker}, G. 2006, \mnras, 369, 1115

\bibitem[{{Sternberg} \& {Soker}(2008)}]{Sternberg_Soker_FatBubbles_2008}
{Sternberg}, A. \& {Soker}, N. 2008, \mnras, 384, 1327

\bibitem[{{Sternberg} \& {Soker}(2009)}]{stemberg_soundwave_2009}
{Sternberg}, A. \& {Soker}, N. 2009, \mnras, 395, 228

\bibitem[{{Storm} {et~al.}(2015){Storm}, {Jeltema}, \&
  {Rudnick}}]{storm_mergingcluster_2015}
{Storm}, E., {Jeltema}, T.~E., \& {Rudnick}, L. 2015, \mnras, 448, 2495

\bibitem[{Sutherland \& Bicknell(2007)}]{sutherland_interactions_2007}
Sutherland, R.~S. \& Bicknell, G.~V. 2007, \apjs, 173, 37

\bibitem[{Sutherland \& Dopita(1993)}]{sutherland_cooling_1993}
Sutherland, R.~S. \& Dopita, M.~A. 1993, Astrophysical Journal Supplement
  Series (ISSN 0067-0049), 88, 253

\bibitem[{{Vernaleo} \& {Reynolds}(2006)}]{vernaleo_problems_2006}
{Vernaleo}, J.~C. \& {Reynolds}, C.~S. 2006, \apj, 645, 83

\bibitem[{{Voit} {et~al.}(2017){Voit}, {Meece}, {Li}, {O'Shea}, {Bryan}, \&
  {Donahue}}]{Voit2017}
{Voit}, G.~M., {Meece}, G., {Li}, Y., {et~al.} 2017, \apj, 845, 80

\bibitem[{{Wik} {et~al.}(2014){Wik}, {Hornstrup}, {Molendi}, {Madejski},
  {Harrison}, {Zoglauer}, {Grefenstette}, {Gastaldello}, {Madsen},
  {Westergaard}, {Ferreira}, {Kitaguchi}, {Pedersen}, {Boggs}, {Christensen},
  {Craig}, {Hailey}, {Stern}, \& {Zhang}}]{wik_bulletcluster_2014}
{Wik}, D.~R., {Hornstrup}, A., {Molendi}, S., {et~al.} 2014, \apj, 792, 48

\bibitem[{{Wilman} {et~al.}(2009){Wilman}, {Edge}, \& {Swinbank}}]{Wilman2009}
{Wilman}, R.~J., {Edge}, A.~C., \& {Swinbank}, A.~M. 2009, \mnras, 395, 1355

\bibitem[{{Wilms} {et~al.}(2000){Wilms}, {Allen}, \&
  {McCray}}]{wilms_absorption_2000}
{Wilms}, J., {Allen}, A., \& {McCray}, R. 2000, \apj, 542, 914

\bibitem[{{Wise} {et~al.}(2007){Wise}, {McNamara}, {Nulsen}, {Houck}, \&
  {David}}]{wise_hydraA_2007}
{Wise}, M.~W., {McNamara}, B.~R., {Nulsen}, P.~E.~J., {Houck}, J.~C., \&
  {David}, L.~P. 2007, \apj, 659, 1153

\bibitem[{Yang \& Reynolds(2016)}]{yang_how_2016}
Yang, H.-Y.~K. \& Reynolds, C.~S. 2016, \apj, 829, 90

\bibitem[{Zhuravleva {et~al.}(2016)Zhuravleva, Churazov, Arevalo, Schekochihin,
  Forman, Allen, Simionescu, Sunyaev, Vikhlinin, \&
  Werner}]{zhuravleva_nature_2016}
Zhuravleva, I., Churazov, E., Arevalo, P., {et~al.} 2016, \mnras, 458, 2902

\end{thebibliography}

\end{document}